\documentclass[
    aps,prb,twocolumn,
	groupedaddress,superscriptaddress,
	amsfonts,amssymb,amsmath,
	citeautoscript,longbibliography,
	letterpaper, nofootinbib
	]{revtex4-2}

\usepackage[utf8]{inputenc}
\usepackage[english]{babel}

\usepackage{microtype} 
\usepackage{xspace} 

\usepackage{txfonts}  
\usepackage{txfontsb} 

\usepackage{bm} 

\usepackage{xcolor}
\usepackage[]{graphicx} 
\graphicspath{{figs/}}

\usepackage[]{booktabs}
\usepackage{array}
\usepackage{layouts}
\usepackage{multirow}

\usepackage{enumerate}
\usepackage[inline]{enumitem}

\usepackage{xr}
\makeatletter
\newcommand*{\addFileDependency}[1]{
  \typeout{(#1)}
  \@addtofilelist{#1}
  \IfFileExists{#1}{}{\typeout{No file #1.}}
}
\makeatother

\usepackage{hyperref}
\hypersetup{colorlinks,
	linkcolor={blue!75!black!80!yellow},
	citecolor={blue!75!black!80!yellow},
	urlcolor={blue!75!black!80!yellow}
}

\usepackage[capitalize,nameinlink]{cleveref}

\crefname{subequations}{Eqs.}{Eqs.} 
\Crefname{subequations}{Eqs.}{Eqs.}
\crefformat{subequations}{#2Eqs.~(#1)#3}
\Crefformat{subequations}{#2Eqs.~(#1)#3}
\crefname{page}{p.}{p.} 
\crefname{table}{Table}{Tables}
\crefname{figure}{Figure}{Figures}
\crefname{section}{Section}{Sections}

\usepackage{placeins}

\usepackage{siunitx}
\DeclareSIUnit[number-unit-product = ]\percent{\char`\%} 

\usepackage[centering,hmargin=18mm,tmargin=29.4mm,bmargin=24mm]{geometry}

\usepackage{soul}

\usepackage{textcomp} 
\usepackage{xifthen}
\usepackage{etoolbox}
\newboolean{togglecomments}
\newboolean{toggletodos}
\newboolean{togglechanges}

\setboolean{togglecomments}{true}
\setboolean{toggletodos}{true}
\setboolean{togglechanges}{false} 

\newcommand{\textblacksquare}{$\blacksquare$}
\newcommand{\todo}[1]{\ifbool{toggletodos}%
	{\textcolor{green!60!black}{\small\textsf{{}\textsuperscript{\textsc{\textsf{todo}}}}[\ignorespaces#1]}} 
	{}}     
\newcommand{\comment}[2]{\ifbool{togglecomments}%
		{\textcolor{blue!70!black}{\small\sf\textsuperscript{\textsc{\textsf{\ignorespaces#1}}}[\ignorespaces#2]}} 
		{}}     

\newcommand{\edit}[1]%
		{\textcolor{Black}{\ignorespaces#1}} 
		{}

\newcommand{\reply}[2]{\ifbool{togglecomments}%
		{\textcolor{red!70!black}{\small\sf\textsuperscript{\textsc{\textsf{\ignorespaces#1}}}[\ignorespaces#2]}} 
		{}} 
        
\newcommand{\swap}[2]{\ifbool{togglechanges}
	{\ignorespaces#2}  
	{\textcolor{red!70!black}{[\ignorespaces#1]}\textrightarrow{}\textcolor{green!50!black}{[\ignorespaces#2]}}}
\newcommand{\remove}[1]{\ifbool{togglechanges}
	{}    
	{\textcolor{red!70!black}{\ignorespaces#1}}}
\newcommand{\inset}[1]{\ifbool{togglechanges}
	{\ignorespaces#1}  
	{\textcolor{green!50!black}{\ignorespaces#1}}}

\newcommand{\citeremind}[1]{%
	[\textcolor{blue!75!black!80!yellow}{\textblacksquare%
		\ifthenelse{\isempty{#1}}{}{\textsuperscript{\tiny\textsf{\ignorespaces#1}}}%
	}]\xspace}

\newcommand{\eg}{e.g.,\@\xspace}

\newcommand{\appropto}{\mathrel{\vcenter{
			\offinterlineskip\halign{\hfil$##$\cr
				\propto\cr\noalign{\kern.2pt}\sim\cr\noalign{\kern-2.5pt}}}}}

\DeclareFontFamily{U}{mathx}{\hyphenchar\font45}
\DeclareFontShape{U}{mathx}{m}{n}{<5> <6> <7> <8> <9> <10>
                                  <10.95> <12> <14.4> <17.28> <20.74> <24.88>
                                  mathx10}{}
\DeclareSymbolFont{mathx}{U}{mathx}{m}{n}
\DeclareFontSubstitution{U}{mathx}{m}{n}

\makeatletter
\newcommand{\raisemath}[1]{\mathpalette{\raisem@th{#1}}}
\newcommand{\raisem@th}[3]{\raisebox{#1}{$#2#3$}}
\makeatother

\renewcommand{\paragraph}[1]{\vskip 1ex\noindent\textbf{#1.}~}

\usepackage{braket}
\usepackage[eulergreek]{sansmath}
\makeatletter
\renewcommand\@make@capt@title[2]{%
    \@ifx@empty\float@link{\@firstofone}{\expandafter\href\expandafter{\float@link}}%
    \sisetup{math-sf=\textsf}%
    \sansmath\sffamily\textbf{#1\@caption@fignum@sep}#2 
}%

\makeatother

\graphicspath{{figures/}}
\setboolean{togglecomments}{false}
\setboolean{toggletodos}{true}
\setboolean{togglechanges}{false} 
\usepackage[dvipsnames]{xcolor}

\begin{document}
\title{Gradient-based inverse design of quantum phases}

\author{André Grossi Fonseca$^\bigstar$}
\email{agfons@mit.edu}
\affiliation{Department of Physics, Massachusetts Institute of Technology, Cambridge, Massachusetts 02139, USA}
\affiliation{The NSF Institute for Artificial Intelligence and Fundamental Interactions}

\author{Eric Wang$^\bigstar$}
\affiliation{Department of Physics, Massachusetts Institute of Technology, Cambridge, Massachusetts 02139, USA}

\author{Sachin Vaidya}
\affiliation{Department of Physics, Massachusetts Institute of Technology, Cambridge, Massachusetts 02139, USA}
\affiliation{The NSF Institute for Artificial Intelligence and Fundamental Interactions}
\affiliation{Research Laboratory of Electronics, Massachusetts Institute of Technology, Cambridge, Massachusetts 02139, USA}

\author{Patrick J. Ledwith}
\affiliation{Department of Physics, Harvard University, Cambridge, Massachusetts 02138, USA\\
$^\bigstar$ denotes equal contribution}
\affiliation{Department of Physics, Massachusetts Institute of Technology, Cambridge, Massachusetts 02139, USA}

\author{Ashvin Vishwanath}
\affiliation{Department of Physics, Harvard University, Cambridge, Massachusetts 02138, USA\\
$^\bigstar$ denotes equal contribution}

\author{Marin Solja\v ci\'c}
\affiliation{Department of Physics, Massachusetts Institute of Technology, Cambridge, Massachusetts 02139, USA}
\affiliation{The NSF Institute for Artificial Intelligence and Fundamental Interactions}
\affiliation{Research Laboratory of Electronics, Massachusetts Institute of Technology, Cambridge, Massachusetts 02139, USA}

\begin{abstract}
The discovery and understanding of new quantum phases has time and again transformed both fundamental physics and technology, yet progress often relies on slow, intuition-based theoretical considerations or experimental serendipity. Here, we introduce \edit{``target-phase optimization'',} a general gradient-based framework for phase \edit{search}. We define a function, dubbed ``target-phase loss function'', which encodes fingerprints of a quantum state, thereby recasting \edit{the inverse design of a }phase as a tractable optimization problem in Hamiltonian space. The method is broadly applicable to a wide range of phases and can be interfaced with \edit{different} many-body numerical solvers, \edit{as we explicitly illustrate for topologically ordered states, charge-density waves and Fermi liquids}. As \edit{our main} demonstrations, we apply it to spinless fermions on the kagome lattice using exact diagonalization and discover two distinctive fractional Chern insulators (FCIs): (i) at filling $\nu = 1/3$, a ``non-ideal'' Abelian FCI \edit{that remains stable despite strong deviations from ideal quantum geometry}; and (ii) at $\nu = 1/2$, a non-Abelian FCI stabilized purely by finite-range two-body interactions. These results provide explicit realizations of such types of FCIs and establish a versatile paradigm for systematic quantum-phase \edit{inverse design}.
\end{abstract}
\maketitle 

\section{Introduction}

The study of quantum matter has revealed a remarkable range of universal phenomena that emerge from simple microscopic rules. 
Historically, progress in this field has given rise to both conceptual and technological breakthroughs: the development of band theory enabled the classification of metals and insulators and laid the foundation for the semiconductor industry, while the discovery of superconductivity led to applications such as high-field magnets, widely used in medical imaging and particle accelerators.
More recently, there has been great interest in topologically ordered phases~\cite{wen2004quantum}, whose fractionalized excitations~\cite{laughlin_anomalous_1983} and long-range entanglement~\cite{chen2010entanglement} provide both new insights into the behavior of many-body systems and potential applications in quantum information~\cite{kitaev2003anyons, nayak2008quantum}. 
These efforts place the field at a critical juncture, where new tools are urgently needed to accelerate discovery.

The past few years have also seen a rapid rise of data-driven approaches in the quantum sciences, including machine-learning~\cite{carleo2019ML, Dawid_2025} and optimization-based~\cite{liao2019tn, zhang2023admc} techniques, which promise to transform how complex many-body problems are tackled. 
While such techniques traditionally rely on large, high-quality datasets, recent advances have shown that they can be effective even in scenarios with limited computational or experimental resources, opening the door for a more structured exploration of quantum matter.

Despite great progress, identifying novel quantum phases remains one of the central challenges in many-body physics. 
Strong interactions typically preclude perturbative methods, while brute-force numerical approaches are restricted by the exponential complexity of the problem. 
These difficulties become particularly acute when attempting to map phase diagrams in high-dimensional parameter spaces, where subtle energetic considerations determine the ground state. 
Carefully charting these phase diagrams provides the means to understand how different phases compete and to identify which experimental controls can be tuned to realize and stabilize states of interest. 
However, the absence of general, targeted search protocols has left the \edit{exploration} of novel phases largely to chance. 

\begin{figure*}
    \centering
    \includegraphics[scale=1]
    {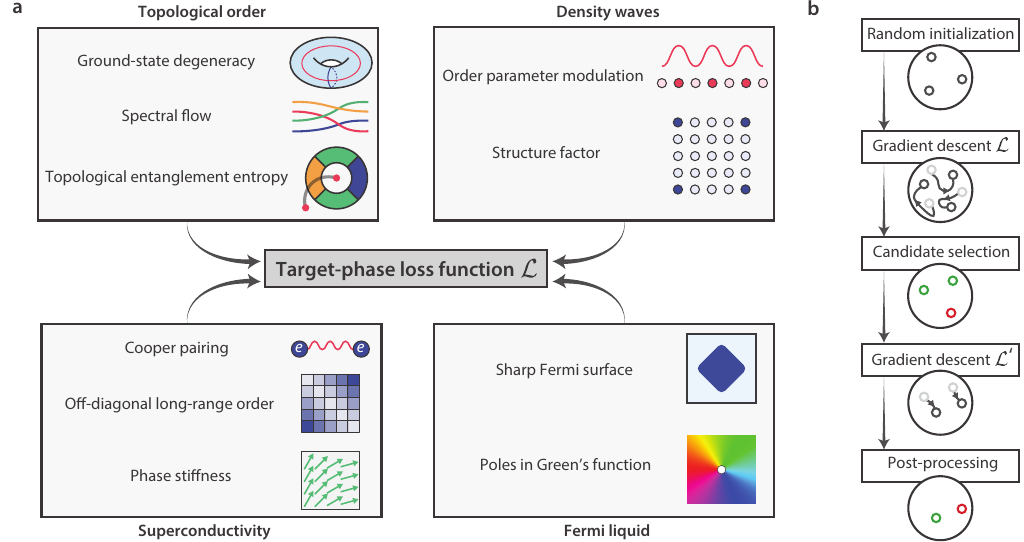}
    \caption{%
     (a)~Diagram illustrating the general procedure for targeting phases of interest.
    A quantum ground state (\eg a topologically ordered state, density wave, superconductor, or Fermi liquid) is characterized by a collection of sharp features (boxed), some of which can be leveraged to define a target-phase loss function $\mathcal{L}$. Optimizing $\mathcal{L}$ over a family of Hamiltonians leads to the identification of microscopic models whose ground states realize the phase of interest.
    (b)~Schematic workflow for gradient-based search for a target phase (see text).}
    \label{fig:figure1}
\end{figure*}

In this work, we address this limitation by proposing a simple yet powerful method, \edit{which we dub ``target-phase optimization''. Its core idea is to }recast the search for quantum ground states across families of Hamiltonians as an optimization problem. 
Central to our approach is the definition of a ``target-phase loss function'', designed to encode features of the desired phase, which can be defined for a wide range of quantum orders (\cref{fig:figure1}a).  
The algorithm relies on a many-body numerical solver of choice to iteratively compute the loss function and explores the parameter space by using gradient descent to minimize the loss, thereby converging onto regions in which the target phase is stabilized.
Our approach is highly modular, allowing for the inclusion of multiple constraints to tailor the search to specific physical criteria. 
In this way, when interfaced with numerical methods for ground-state calculation, our algorithm provides a broadly applicable framework for \edit{inverse-designing} quantum many-body phases.

Here, we focus on phases with ground-state degeneracy, which may arise due to symmetry breaking~\cite{girvin2019modern} or topological order~\cite{wen1990topological},  to formulate an explicit loss in terms of spectral features such as the expected ground-state symmetry sectors~\cite{ bernevig_emergent_2012, wietek2017studying} and spectral flow under adiabatic flux insertion~\cite{laughlin1981flux, oshikawa_commens_2000}.
As exemplary applications of our method, we employ it to search for Abelian and non-Abelian fractional Chern insulators (FCIs) in novel settings, by combining exact diagonalization (ED) with our optimization scheme.
We first focus on discovering ``non-ideal'' Abelian FCIs, which depart from the conventional paradigm of replicating certain geometric features of the lowest Landau level~\cite{Ledwith2020, Wang2021,ledwith2022family, ledwith_vortexability_2023}. 
While previous large-scale searches only found near-ideal FCIs~\cite{jackson_geometric_2015}, we use our targeted search method to find FCIs in \edit{a band with strong deviations from ideal quantum geometry}.
We further delineate the nearby phase diagram, shedding light on competing phases.
Secondly, we apply our method to non-Abelian FCIs, examples of which have so far relied on multi-body or long-range interactions.
We find robust examples of non-Abelian Moore--Read states stabilized by finite-range two-body interactions, which depend non-monotonically on distance, providing a minimal lattice setting in which non-Abelian fractionalization takes place.
These novel FCIs \edit{go beyond} conventional expectations and call for new theories on how fractionalization emerges in these settings.
\edit{To demonstrate the generality of our method, in the Supplemental Material (SM) we provide further examples of target-phase optimization in the context of charge density waves and Fermi liquids.}

Taken together, our results highlight two distinct advances: first, a versatile and general optimization framework for the systematic \edit{inverse design} of quantum phases, and second, its concrete demonstration in uncovering two kinds of FCIs beyond the current theoretical understanding. 
The generality of the method provides a practical route to the \edit{exploration} of a variety of quantum phases across diverse settings.

\section{Target-phase loss function}

\subsection{General considerations and workflow}

Quantum phases of matter are broadly characterized by sharp features encoded in their ground states. 
The precise identification of such features is intimately tied to the definition and discovery of novel phases, both theoretically and experimentally. 
Across the landscape of possible ground states, these features can be remarkably rich, varying widely in both character and physical origin. 
\cref{fig:figure1}a provides an overview of several representative classes of quantum phases and a non-exhaustive list of the features that define them.

Historically, progress in the field has often hinged on the construction of physically viable microscopic models that realize novel phases.
Such a construction supplements general theoretical considerations, in that it offers a concrete setting to study the essential ingredients required for the phase's existence and stability, as well as competing orders. 
Motivated by this, here we focus on the \edit{\emph{inverse design}} question: given a target quantum phase, how can one efficiently identify microscopic Hamiltonians that realize it? 
Our approach, \edit{which we dub ``target-phase optimization'',} is to introduce a continuous function, the ``target-phase loss function'' $\mathcal{L}(\mathbf{p})$, defined in the parameter space $\mathbf{p}$ of a chosen family of models\footnote{Such parameters should be understood in the broadest sense, such as characterizing energetics (\eg hopping amplitudes and interaction strengths), the lattice geometry (\eg lattice vector sizes and angles), external fields, etc.}.
This function is designed to incorporate some of the defining sharp features of the phase, in such a way that the global minima of $\mathcal{L}(\mathbf{p})$ are candidate points for the desired ground state. 
As exemplified below and in the SM, \edit{this construction is quite modular: any phase for which a suitable differentiable loss function can be constructed and computed through the chosen numerical solver can, in principle, be targeted.}
In doing this, we effectively recast general phase search as an optimization problem, which can be tackled by a vast number of numerical techniques, most prominently gradient descent. 

Our strategy offers several conceptual and practical advantages.
Unlike random or grid‐based sampling, which require a number of expensive function evaluations that grows exponentially with parameter space dimensionality, methods such as gradient descent navigate parameter space with informed directional updates and through one-dimensional (1D) paths, dramatically reducing the number of calculations and providing a more favorable scaling.
Furthermore, a gradient-based approach enables the use of advanced numerical techniques, including modern optimizers~\cite{kingma_adam_2017} and automatic differentiation~\cite{baydin_automatic_2018}, to accelerate convergence.
The modularity of the loss function also opens avenues for targeting a wide range of exotic quantum phases by embedding appropriate physical fingerprints. 
As a result, \edit{target-phase optimization} provides a flexible framework that complements existing numerical techniques, particularly in regimes where exhaustive exploration of the parameter space is computationally prohibitive.

Given some target-phase loss function, we now describe our search workflow in detail, schematically shown in \cref{fig:figure1}b. 
We begin by choosing a family of microscopic Hamiltonians residing in some parameter space $\mathbf{p}$, and a system size with which to perform the searches.
Initial parameter seeds are then generated within a predetermined range. 
For each such seed, we iteratively minimize $\mathcal{L}(\mathbf{p})$ until either the loss becomes sufficiently low, indicating a candidate point for the target phase, or a local minimum is reached.
Candidate points are collected for further analysis; if none appear, one returns to the first step and modifies initial choices, such as enlarging the sampling range, introducing additional parameters, or employing a different Hamiltonian family. 
An optional next step is to define a refinement loss function $\mathcal{L}'(\mathbf{p})$, tailored to optimize specific properties of the phase, and explore the local regions of candidate points via gradient descent. 
Finally, we confirm the existence of the desired phase through rigorous post‐processing, such as scaling up the system size, further converging numerical calculations and computing additional complementary features of the ground state. 

\subsection{Example: ground-state degeneracy in exact diagonalization}

\begin{figure}
    \centering
    \includegraphics[scale=1]
    {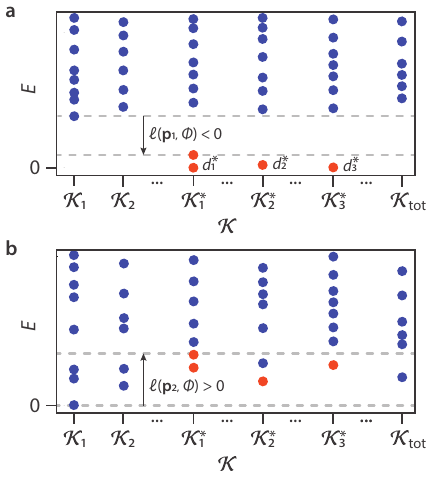}
    \caption{%
    Generic spectra of a many-body Hamiltonian at two distinct parameter points (a) $\mathbf{p}_1$ and (b) $\mathbf{p}_2$ and a given flux $\Phi$, with energy $E$ as a function of symmetry sectors $\mathcal{K}$.
    The target and complement manifolds are shown in orange and blue, respectively.
    The per-flux loss $\ell(\mathbf{p}_1, \Phi)$ is negative, indicating that the system is potentially in the target phase with many-body gap $-\ell(\mathbf{p}_1, \Phi)$, whereas $\ell(\mathbf{p}_2, \Phi) > 0$ captures a spectral distance to the target phase in parameter space.
    }
    \label{fig:figure2_new}
\end{figure}

We now provide an explicit example for a target-phase loss function suitable for phases with ground-state degeneracy on a torus, which may arise due to symmetry breaking (\eg charge and spin density waves~\cite{gruner2018density}) or topological order (\eg fractional quantum Hall states~\cite{tsui_fqhe_1982, laughlin_anomalous_1983} and quantum spin liquids~\cite{savary_quantum_2017}).
A general many-body Hamiltonian $H(\mathbf{p})$ often possesses symmetries, implying that it can be cast into a block‐diagonal form in a symmetry-adapted basis, where each block corresponds to a set of conserved quantum numbers $\mathcal{K}_i$, such as charge, momentum, or spin.
Therefore, the spectrum of such a many-body system can be organized in different symmetry sectors $\{ \boldsymbol{E}_{\mathcal{K}_i} (\mathbf{p}) \}$ (\cref{fig:figure2_new}). 
For example, in the ED method~\cite{jafari_introduction_2008}, one has direct access to $\{ \boldsymbol{E}_{\mathcal{K}_i} (\mathbf{p}) \}$, as well as corresponding eigenstates, by diagonalization of the blocks in each $\mathcal{K}_i$ sector separately. 
For our purposes, it is also useful to include a threaded magnetic flux through the torus $\Phi$ as an additional parameter, which is equivalent to twisting the boundary conditions. 
This is because, for a gapped phase, inserting $2\pi$ flux corresponds to a unitary operation on the ground-state manifold~\cite{laughlin1981flux, oshikawa_commens_2000}, and therefore the gap should be robust to adiabatic flux insertion.
Thus, the flux-resolved low-energy spectrum $\{ \boldsymbol{E}_{\mathcal{K}_i} (\mathbf{p}, \Phi) \}$ constitutes the primary data from which we construct the loss function introduced below.
Because such data can currently be computed reliably only in ED, in the SM we discuss target-phase loss functions that do not rely on the energy spectrum, and show explicit examples of a gradient-based search for \edit{both} charge-density waves \edit{and Fermi liquids, }employing both ED and auxiliary-field quantum Monte Carlo.

In ED, due to finite-size effects, the ground-state degeneracy typically appears as quasi-degenerate states, with the ground-state manifold located at a set of symmetry sectors $\{\mathcal{K}_i^*\}$, with degeneracies $\{d_i^*\}$; \edit{note that the symmetry sectors and degeneracies we may use are completely general}.
In many cases, the data $\{(\mathcal{K}_i^*, d_i^*)\}$ for a particular degenerate quantum phase is known a priori from group-theoretic~\cite{wietek2017studying} or topological~\cite{regnault_fractional_2011, bernevig_emergent_2012, ardonne2008degeneracy, lu_filling-enforced_2020, cheng_ordering_2025,qispinliquid2015} analysis\edit{, or, if available, by comparison to a parent Hamiltonian for which the ground state is known to be in the phase of interest}.
Then, to define the loss we first partition the low-energy spectrum into a \emph{target manifold}, consisting of the lowest $d_i^*$ levels in each $\mathcal{K}_i^*$ sector, and a \emph{complement manifold} comprising all remaining levels. 
Importantly, this partitioning remains well-defined even when the system does not realize the target quantum phase. 
The objective of the optimization is to promote the target manifold to become the true ground-state manifold. 
To do so, we introduce the following quantity
\begin{equation}
\label{eq:loss_per_flux}
    \ell(\mathbf{p}, \Phi) = \max_{E \in\, \operatorname{target}}E(\mathbf{p}, \Phi)  - \min_{E \in\, \operatorname{complement}}E(\mathbf{p}, \Phi), 
\end{equation}

which measures the gap between the highest energy in the target manifold and the lowest energy in the complement manifold at a particular $\Phi$.
A negative value $\ell(\mathbf{p}, \Phi) < 0$ indicates that all levels in the target manifold lie below any competing states, yielding a candidate point for the target phase at $\mathbf{p}$ with associated many-body gap $\Delta_\Phi(\mathbf{p}) = -\ell(\mathbf{p}, \Phi)$. 
If instead $\ell(\mathbf{p}, \Phi) > 0$, this quantity encodes a spectral distance from the target quantum phase, providing a concrete measure for quantifying deviations from the desired ground state (\cref{fig:figure2_new}). 
Finally, to ensure stability of this gap under flux insertion, we define the target-phase loss function as
\begin{equation}
\label{eq:general_loss}
    \mathcal{L}(\mathbf{p}) = \max_{\{ \Phi_j \}} \ell(\mathbf{p}, \Phi_j)
\end{equation}
where the maximum is taken over a user-defined finite sampling of flux values $\{ \Phi_j \}$. 
By construction, $\mathcal{L}(\mathbf{p}) < 0$ if and only if $\ell(\mathbf{p}, \Phi_j) < 0$ for every $\Phi_j$, guaranteeing that the ground‐state degeneracy and corresponding gap persist under the sampled fluxes. 
Although the $\min$ and $\max$ functions  introduce non-differentiable points in the loss landscape, $\mathcal{L}(\mathbf{p})$ remains piecewise differentiable almost everywhere.
This enables the use of gradient‐based methods to find regions of parameter space for which $\mathcal{L}(\mathbf{p}) < 0$\footnote{In general, it may be beneficial to also keep points with marginally positive loss in this step. This is because the searches are done at modest system sizes, which may be smaller than the phase's correlation length, so scaling up can lead to sizable gaps even if there is none at small clusters.}.
Note that other physical properties of interest can be encoded as additional arguments of the $\max$ function in \cref{eq:general_loss}, underscoring its modularity.

\section{Non-ideal Abelian fractional Chern insulators}

We now apply \edit{target-phase optimization} to search for examples of non-ideal FCIs.
FCIs are lattice analogs of fractional quantum Hall states, with time-reversal broken either spontaneously or through an external magnetic field~\cite{parameswaran2013fractional,bergholtz_topological_2013,liu_recent_2024,neupert_fci_2011,sun_2011_flatband, tang_high-temperature_2011,Sheng2011,regnault_fractional_2011, barkeshli2012nematic}. 
Compelling numerical evidence has established the existence of both fermionic~\cite{neupert_fci_2011,Sheng2011,regnault_fractional_2011, wu_zoology_2012} and bosonic~\cite{hafezi_fractional_2007, wang_fractional_2011} FCIs in tight‐binding models with strong interactions at partial fillings of flat topological bands. 
More recently, theoretical predictions~\cite{Senthil_NearlyFlatBand,Ledwith2020,ZhaoLiu_TBG,Cecile_TBG_Flatband,Lanchli21,Valentin22_anomaloushallmetal,Kaisun_FCI21,MoralesDuranTMDFCI} and experimental realizations of FCIs in moir\'e materials~\cite{spanton2018observation, xie2021fractional,park_observation_2023, zeng2023thermodynamic,lu_fractional_2024}, as well as a two-particle Laughlin state in ultracold atoms~\cite{leonard2023realization}, have reignited interest in engineering lattice Hamiltonians that support fractionalized phases.

\begin{figure*}
    \centering
    \includegraphics[scale=1]
    {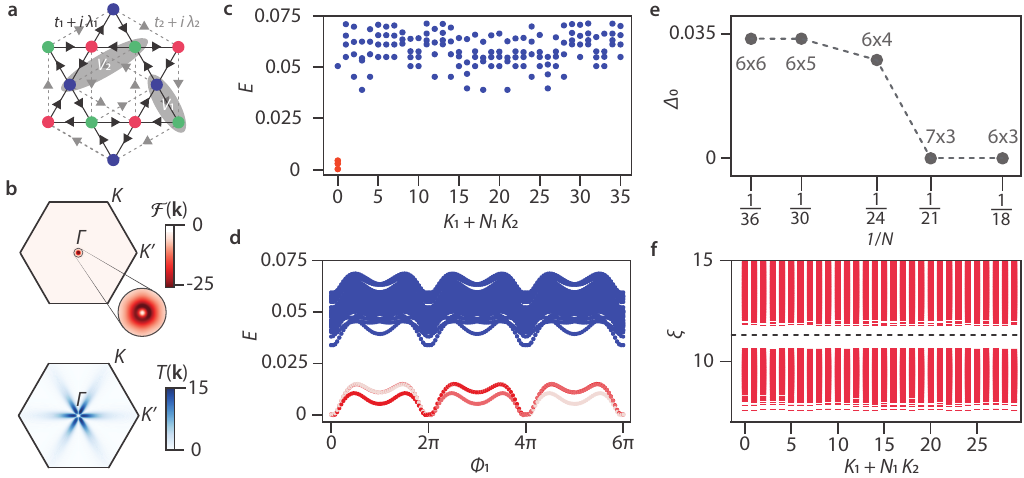}
    \caption{%
    ED evidence for a non-ideal FCI, located at $(\lambda_1, t_2, \lambda_2, V_2) = (-0.169, -0.992, -0.157, 0.25)$.
    (a)~The kagome lattice model with first- and second-neighbor complex hoppings, as well as first- and second-neighbor density-density interactions. 
    (b)~Single-particle Berry curvature $\mathcal{F}(\mathbf{k})$ (top) and momentum-resolved trace violation $T(\mathbf{k})$ (bottom) distributions in the Brillouin zone.
    High-symmetry momenta $\Gamma = (0, 0), K = (2\pi/3, 2\pi/\sqrt{3}), K' = (4\pi/3, 0)$ are labeled.
    Inset: zoomed in Berry curvature distribution near the $\Gamma$ point, showing that it vanishes at $\Gamma$, consistent with a quadratic band touching with small gap.
    (c)~Zero-flux ED spectrum for a $6 \times 6$ cluster as a function of center-of-mass momentum at filling $\nu=1/3$.
    (d)~ED spectrum for a $6 \times 5$ cluster as a function of magnetic flux $\Phi_1$ along reciprocal lattice vector $\mathbf{b}_1$.
    (e)~Zero-flux gap $\Delta_0$ as a function of inverse system size $1/N$.
    All clusters with $N < 24$ have vanishing gap. 
    (f)~Particle entanglement spectrum on a $6 \times 5$ cluster for subspace particle number $N_A = 4$.
    The number of states below the dashed line is 9975, consistent with a Laughlin ground state.
    }
    \label{fig:figure2}
\end{figure*}

Throughout these developments, the conditions under which FCIs, particularly Laughlin-like states, are stabilized have been intensely investigated. 
The most important set of conditions is related to band geometry~\cite{parameswaran2013fractional,liu_recent_2024,roy_band_2014,jackson_geometric_2015,Lee2017,Ledwith2020,ozawa_relations_2021,Mera2021b,Wang2021,varjas_topological_2019,ledwith_vortexability_2023,Estienne2023,Okuma}.
Concretely, the quantum geometry of a band is contained in the quantum geometric tensor~\cite{parameswaran2013fractional,liu_recent_2024}
\begin{equation}
\label{eq:qgt}
    \eta^{\mu\nu}(\mathbf{k}) = \sum_n \bra{\partial_{k_\mu}u_n(\mathbf{k})}Q(\mathbf{k})\ket{\partial_{k_\nu}u_n(\mathbf{k})}, 
\end{equation}
where $u_n(\mathbf{k})$ is a cell-periodic Bloch state in band $n$ with momentum $\mathbf{k}$, and $Q(\mathbf{k}) = 1 - \sum_n \ket{u_n(\mathbf{k})}\bra{u_n(\mathbf{k})}$.
Its real and imaginary parts are the Fubini--Study metric and Berry curvature, respectively:
\begin{equation}
\label{eq:fsmetric+bc}
    g^{\mu\nu}(\mathbf{k}) = \operatorname{Re} \eta^{\mu\nu}(\mathbf{k}), \quad \mathcal{F}(\mathbf{k}) \varepsilon^{\mu\nu} = \frac{1}{2} \operatorname{Im} \eta^{\mu\nu}(\mathbf{k}),
\end{equation}
with $\varepsilon^{\mu\nu}$ the 2D Levi--Civita symbol. 

Geometric similarity to LLs is then quantified by two quantities, the ``trace violation'' $T$ and Berry curvature fluctuations $\sigma_B$, defined as

\begin{equation}
\label{eq:spmetrics}
\begin{split}
    &T = \frac{1}{2\pi} \int_{\operatorname{BZ}} d^2k\, T(\mathbf{k}), \quad T(\mathbf{k}) = \operatorname{Tr}g^{\mu\nu}(\mathbf{k}) - |\mathcal{F}(\mathbf{k})|, \\   
    &\sigma_B = \sqrt{\int_{\operatorname{BZ}} \frac{d^2k}{A_{BZ}}\left(\frac{\mathcal{F}(\mathbf{k})}{\mathcal{F}_0} - C\right)^2\,}.
\end{split}
\end{equation}
Here $A_{BZ}$ is the Brillouin zone (BZ) area, $\mathcal{F}_0 = 2\pi/A_{BZ}$ and $C$ the Chern number, with normalization chosen such that both $T$ and $\sigma_B$ are dimensionless. 
Such quantities can be calculated for the $n$-th LL to be $T=2n$ and $\sigma_B=0$~\cite{ozawa_relations_2021}. 

Initially, these band-geometric criteria were applied to FCIs for the purpose of perfectly mimicking LL physics~\cite{parameswaran2013fractional,roy_band_2014, jackson_geometric_2015,Lee2017}. Indeed, if a flat band with density-density interactions has quantum geometry such that $T = \sigma_B = 0$, then the projected density operators and resulting interacting physics will be identical to those of the lowest LL \cite{roy_band_2014}. 
More recently, it has been understood that the lowest LL does not need to be fully mimicked in order to guarantee FCI ground states~\cite{Ledwith2020,Wang2021,ledwith2022family,ledwith_vortexability_2023,Dong2023Many,wang2023origin}. 
If a band satisfies the ``trace condition'' $T(\mathbf{k}) = 0$, thereby having ``ideal'' geometry, then its wavefunctions can be multiplied by a vortex function without exciting remote band states~\cite{ledwith2022family,ledwith_vortexability_2023}. 
This condition, known as vortexability, enables the construction of exact FCI ground states of short-range repulsive interactions, in direct analogy to the construction of the Laughlin wavefunction~\cite{pokrovskySimpleModelFractional1985,TrugmanKivelson1985}.
Finally, note that the trace condition can be somewhat relaxed, as both the lowest and first LLs host Laughlin states at filling $\nu=1/3$, but higher Landau levels do not~\cite{foglerStripeBubblePhases2001a}. 
This suggests that bands with $T \lesssim 2$ can be sufficiently ideal and fit well within this band-geometric paradigm. 

\edit{Despite their usefulness, whether these conditions are necessary for FCI stability remains an open question. In particular, a nonzero Chern number together with vortexability, or ideal quantum geometry, is sufficient to stabilize the Laughlin state under appropriate short-range interactions~\cite{ledwith_vortexability_2023}. This raises the question of whether either ingredient is individually essential. Recent work has shown that FCIs can arise in topologically trivial isolated bands~\cite{lin_fractional_2025}, although these bands retain important features associated with ideal bands. Other studies in tight-binding~\cite{simon_fractional_2015, yang_singular_2025} and continuum moiré models~\cite{dong_theory_2024} provide further clues about the ingredients required for FCI stability, but do not directly test whether band ideality itself is essential.}

\edit{Here, we address this using our gradient-based optimization method to uncover explicit examples of non-ideal FCIs, which we define as those with trace violations closer to higher LLs ($T>3$).} Our example will also feature strongly fluctuating Berry curvature $\sigma_B \gg 1$, making it distinct from any LL.
To this end, we focus on spinless fermions at $\nu=1/3$ on the kagome lattice. 
The absence of spin means that, for a finite cluster on a torus geometry with $N_1 \times N_2$ unit cells, the symmetry sectors are the $N_1 N_2$ allowed values of the many-body crystal momentum, which we choose to be $\mathcal{K}_i = (K_x, K_y) = \tfrac{K_1}{N_1} \mathbf{b}_1 + \tfrac{K_2}{N_2} \mathbf{b}_2$, with integers $0 \leq K_1 < N_1, 0 \leq K_2 < N_2$ and $\mathbf{b}_{1,2}$ primitive reciprocal lattice vectors.
The Hamiltonian reads
\begin{equation}
\begin{split}
&H = H_{\mathrm{kin}} + H_{\mathrm{int}}, \\
&H_{\mathrm{kin}} = -\sum_{\langle i,j\rangle} (t_{1}+i\nu_{ij}\lambda_1) c_{i}^{\dagger}c^{\phantom{}}_{j} 
- \sum_{\langle\langle i,j\rangle\rangle}(t_{2}+i\nu_{ij}\lambda_2)c_{i}^{\dagger}c^{\phantom{}}_{j}, \\
&H_{\mathrm{int}} = V_{1}\sum_{\langle i,j\rangle}:\!n^{\phantom{}}_{i}\,n^{\phantom{}}_{j}\!: + V_{2}\sum_{\langle\langle i,j\rangle\rangle}:\!n^{\phantom{}}_{i}\,n^{\phantom{}}_{j}\!:,
\end{split}
\end{equation}
where $c^{\phantom{}}_i\, (c_i^\dag)$ is the fermion annihilation (creation) operator at site $i$, $n^{\phantom{}}_i = c_i^\dag c^{\phantom{}}_i$ is the density operator, and $:\,\,:$ indicates normal ordering.
The first- and second-neighbor sites are represented by $\langle i,j\rangle, \langle\langle i,j\rangle\rangle$.
$t_1$ and $\lambda_1$ parameterize the real and imaginary parts of the first-neighbor hopping, $t_2$ and $\lambda_2$ parameterize the real and imaginary parts of the second-neighbor hopping, and $\nu_{ij}=\pm 1$ encodes a chosen chirality for complex hoppings, as shown in \cref{fig:figure2}a.
Interactions are included through two-body first- and second-neighbor density-density terms with strength $V_1$ and $V_2$, respectively.
Throughout, we set $V_1=1$ to fix the energy scale. 

Because our interest lies in the quantum geometry of a single isolated band, we first flatten the lowest single-particle band and project the ED calculation to it~\cite{bergholtz_topological_2013, regnault_fractional_2011}, which effectively amounts to sending single‐particle gaps to infinity without altering geometric properties. 
In this projected regime, only the single‐particle eigenstates enter the many‐body calculation, so without loss of generality one may set $t_1 = 1$ as well.
Consequently, the effective parameter space is four-dimensional (4D), $\mathbf{p}=(\lambda_{1},t_{2},\lambda_{2},V_{2})$.
In this regime, the points around $\mathbf{p} =(1,0,0,0)$ have been shown to be part of an ideal FCI region, exhibiting small $T$ and $\sigma_{B}$~\cite{wu_zoology_2012, jackson_geometric_2015}.

We initiate the target-phase optimization on a $N_1 \times N_2 =  3\times 3$ cluster, which for the kagome lattice is already large enough to be able to resolve the ideal FCI region while \edit{preserving the $C_3$ symmetry of the lattice and including the $K$ and $K'$ points in the Brillouin zone, which is appropriate for resolving competing charge-density waves.
These properties make this small cluster sufficient for our purposes; in the next section, we show that we can employ much larger clusters for the searches when needed}.
At this system size and filling, the three FCI ground states are expected from the generalized Pauli principle to be all located at $\mathcal{K}^* = (0, 0)$, with associated degeneracy $d^* = 3$~\cite{regnault_fractional_2011, bernevig_emergent_2012}, which defines the target and complement manifolds.
We choose the extremal flux values $\{ \Phi \} = \{0, \pi \}$ for calculating the target-phase loss function.
Employing the workflow described above, we first run gradient descent on the target-phase loss function defined in \eqref{eq:general_loss} to locate FCI regions, and then define a refinement loss $\mathcal{L}'(\mathbf{p})=-T(\mathbf{p})$ to maximize the trace violation of promising candidates. 
Gradient descent is carried out using the Adam optimizer~\cite{kingma_adam_2017}, using learning rate $0.1$ in the initial search to ensure thorough exploration of parameter space, and learning rate $0.001$ in the refinement step for a more fine-grained search within FCI regions. 
Gradients of the loss functions were evaluated via automatic differentiation of the ED output for computational efficiency~\cite{baydin_automatic_2018}.

The outcome of this targeted search is illustrated in \cref{fig:figure2}. 
We identify a non‐ideal FCI phase in a $C = -1$ band at $\mathbf{p} = (\lambda_{1},t_{2},\lambda_{2},V_{2}) = (-0.169, -0.992, -0.157, 0.25)$  characterized by $T \approx 3.3$ and $\sigma_{B} \approx 5.2$. 
Plots of $\mathcal{F}(\mathbf{k})$ and $T(\mathbf{k})$ (\cref{fig:figure2}a,b) reveal that both quantities are sharply concentrated near the $\Gamma$ point, in stark contrast to the nearly uniform distributions expected for LL-like FCIs. 
In the SM, we show that the origin of this non-uniform distribution is that the lowest single-particle band is near a quadratic band crossing with the middle band.
However, we emphasize that many-body calculations are performed in the projected limit of the flattened lowest band, which removes band mixing.
\edit{In the SM we consider a generalized ten-dimensional family of time-reversal-broken three-band Hamiltonians. Despite the substantially larger parameter space, the optimization efficiently identifies FCIs with even stronger deviations from ideal quantum geometry, reaching $T \approx 12.08$.}

\begin{figure}
    \centering
    \includegraphics[scale=1]
    {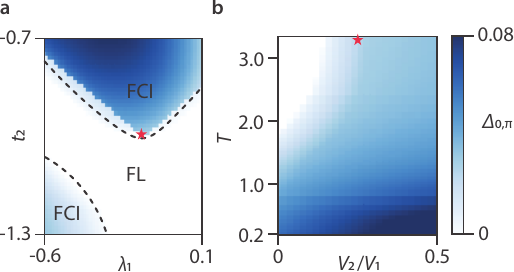}
    \caption{%
    (a)~Phase diagram as a function of $(\lambda_1, t_2)$ at fixed $(\lambda_2, V_2) =(-0.157, 0.25)$ for a $6 \times 4$ cluster. 
    The FCI gap $\Delta_{0, \pi} = \min(\Delta_0, \Delta_\pi)$, combining the gaps at fluxes $\Phi=\{0, \pi\}$, is shown in blue. 
    FCI and Fermi liquid (FL) regions are highlighted, and
    dashed lines are guides for the eye.
    (b)~$\Delta_{0, \pi}$ as a function of trace violation $T$ and the ratio of second-neighbor to first-neighbor interaction strengths $V_2/V_1$ along a 1D path in the FCI region found.
    Red star in (a) and (b) indicates the parameter point hosting the non-ideal FCI investigated here.
    }
    \label{fig:figure3}
\end{figure}

On a $6 \times 6$ cluster, ED calculations show a clear many‐body gap separating a triply quasi‐degenerate ground state manifold from higher excitations, located in momentum sectors in accordance with the generalized Pauli principle~\cite{regnault_fractional_2011, bernevig_emergent_2012} (\cref{fig:figure2}c). 
Flux insertion along $\mathbf{b}_1$ on a $6 \times 5$ cluster shows spectral flow among these three states, which remain energetically isolated from excited states and only return to their initial configuration after $6\pi$ flux insertion (\cref{fig:figure2}d). 
\edit{We also compute the many-body Chern number $\mathcal{C}$ by integrating the many-body Berry curvature over twisted boundary conditions and averaging over the three ground-states \cite{niu1985hall}. 
Using a $40 \times 40$ flux grid on the $6 \times 5$ cluster, we find $\mathcal{C} = 1/3 + 5 \times 10^{-3}$, which together with the spectral flow, implies a Hall conductance $\sigma_{xy} = 1/3 \, e^2/h$, providing further evidence consistent with a gapped ground state with fractional charge.}

We also note pronounced finite‐size effects: a clear FCI gap only emerges for sufficiently large lattices (\cref{fig:figure2}e), \edit{which we discuss further below}.
As a final piece of evidence, we show in \cref{fig:figure2}f the particle-cut entanglement spectrum on the $6 \times 5$ cluster.
This is obtained by partitioning the system with $N_p$ particles into two subspaces, $N_p= N_A + N_B$ and computing the set $\{ \xi = -\log \epsilon_i \}$ where $\epsilon_i$ are the eigenvalues of the
reduced density matrix $\rho_A = \operatorname{Tr}_{N_B} \rho$, with $\rho = \tfrac{1}{3} \sum_{i=1}^3 \ket{\Psi_i}\bra{\Psi_i}$ the density matrix for the three quasi-degenerate ground states.
The particle entanglement spectrum exhibits a visible entanglement gap, the number of states below which is consistent with quasihole counting rules~\cite{regnault_fractional_2011, regnault_entanglement_2015}.
Finally, in the SM we also show that the occupation of the momentum orbitals for the three ground states is nearly uniform, further confirming the FCI ground state.\footnote{We note that this non-ideal FCI point has small positive loss on a $3 \times 3$ cluster due to the refinement step, but the many-body gap becomes large for larger clusters as shown in~\cref{fig:figure2}e.} 

\begin{figure*}[!htpb]
    \centering
    \includegraphics[scale=1]
    {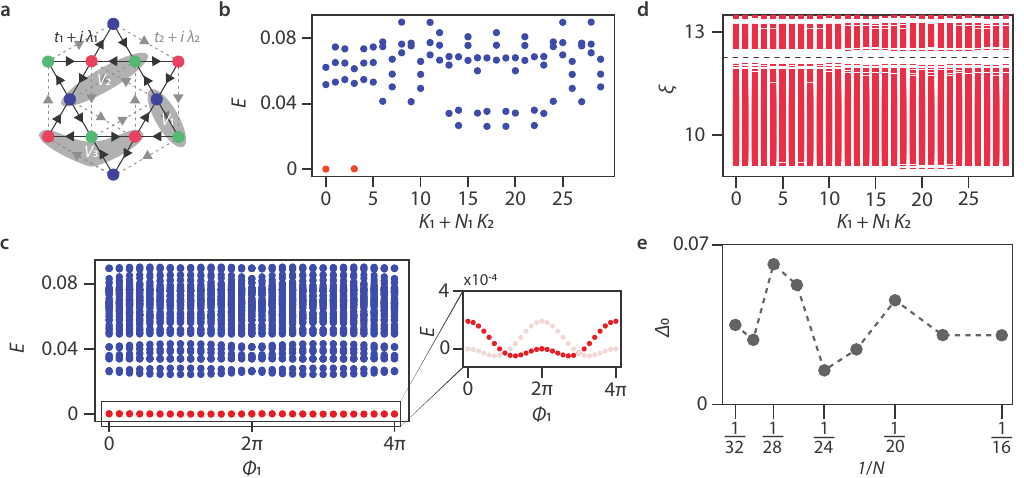}
    \caption{%
    ED evidence for a non-Abelian FCI from finite-range two-body interactions, located at $(\lambda_1, t_2, \lambda_2, V_2, V_3) = (-0.952, 0.956, -0.369, 0.781, 0.987)$.
    (a)~The kagome lattice model with first- and second-neighbor complex hoppings, as well as first-, second- and third-neighbor density-density interactions. 
    (b)~Zero-flux ED spectrum for a $6 \times 5$ cluster as a function of center-of-mass momentum at filling $\nu=1/2$.
    (c)~ED spectrum for a $6 \times 5$ cluster as a function of magnetic flux $\Phi_1$ along reciprocal lattice vector $\mathbf{b}_1$. Inset: zoomed in spectrum showing spectral flow of the ground state manifold.
    (d)~Particle entanglement spectrum on a $6 \times 5$ cluster for subspace particle number $N_A = 4$. 
    The number of states below the dashed line is 25185, consistent with a MR ground state.
    (e)~Zero-flux gap $\Delta_0$ as a function of inverse system size $1/N$.
    }
    \label{fig:figure4}
\end{figure*}

To further probe this unusual phase, we show in \cref{fig:figure3}a the phase diagram in the vicinity of the FCI region, on a 2D slice of the 4D parameter space containing the non-ideal FCI. 
We observe that the main competing phase is a Fermi liquid, which we diagnose via the momentum-space occupation in the SM. 
The absence of charge density waves, an important class of competing phases to FCIs, may be related to the lattice and parameters used, and may help explain the FCI stability in the presence of unusually large $T$ and $\sigma_B$.
\edit{We also observe that the non-ideal FCI point lies quite close to the phase boundary. The pronounced size dependence of its gap may therefore reflect an enhanced correlation length near the transition.
However, we note that such strong finite-size effects or closeness to phase boundaries are not generic to non-ideal FCIs, as illustrated by the generalized three-band model in the SM.}

We also explore how decreasing $T$ within the region affects the FCI stability. 
To do so, we fix $V_{2}$ and run gradient descent with a new refinement loss $\mathcal{L}''(\mathbf{p})=T(\mathbf{p})$, then record the many-body gap along the gradient descent path, and repeat for several values of $V_{2}$. 
As shown in \cref{fig:figure3}b, we observe that when $T$ is tuned toward zero, a FCI is already stabilized at $V_{2} = 0$. 
This is consistent with lowest-LL physics, where short-range interactions favor the fractional quantum Hall effect, as well as the fact that bands with $T=0$ have exact FCI ground states under such interactions~\cite{ledwith_vortexability_2023,wang2023origin}. 
In the SM, we show many-body spectra at selected points along this path, emphasizing that the gap remains open as one tunes parameters from the high-$T$, high-$\sigma_B$ point to the ideal regime.
In addition, we observe that larger values of $T$ require larger values of the longer-range interaction $V_2$ for the FCI to be stabilized.
We conjecture that longer-range interactions together with large $T$ conspire to mimic short-range interactions with small $T$, a version of which can be made precise for the $n$-th LLs under special long-range interactions~\cite{foglerStripeBubblePhases2001a, yoshioka2013quantum}.

\edit{The search we carry out in the SM on the 10D Hamiltonian family also provides a test of the scalability of our approach: target-phase optimization efficiently locates highly non-ideal FCIs and substantially outperforms brute-force searches, which we directly quantify.}

\section{Two-body non-Abelian fractional Chern insulators}

While lattice realizations of Abelian FCIs, particularly Laughlin-like states at fractional fillings, are now well established both theoretically and numerically, far less is known about their non-Abelian counterparts~\cite{moore1991nonabelion, read2000paired}. 
Interest in such phases has grown considerably due to their exotic anyonic quasiparticles, whose non-Abelian braiding statistics offer a platform for fault-tolerant quantum computation~\cite{kitaev2003anyons, nayak2008quantum}. 
Among the most prominent examples is the Moore–Read (MR) state, which supports Ising anyons and has been realized in fractional quantum Hall systems~\cite{willett1987fqhe} and recently suggested to appear in some moiré crystals~\cite{reddy2024moire, ahn2024nonabelian, donna2025nonabelian, liu2024non, xu2025multiple, chong2023higher}. 
However, attempts to realize lattice analogues of the MR state have employed fine-tuned multi-body interactions~\cite{wu_zoology_2012, wang2012nonabelian, stredyniak2013nonabelian, wu2013bloch, bergholtz2015weyl, behrmann2016model} that mimic properties of the continuum MR parent Hamiltonian, or infinite-range two-body interactions~\cite{liu2013longrange, wang2015dipolar}, whose long-range nature complicates determining the thermodynamic ground state from finite-size numerics. 
This raises the question of whether a non-Abelian FCI can be realized in a simple lattice model with only finite-range, two-body interactions. 
This question is of both conceptual and practical significance, since such a minimal model could both allow for exact analytical statements and offer insights into the microscopic ingredients necessary for such topological phases, thereby serving as a guide for experimental realization.

To investigate this possibility, we apply our optimization method to the same kagome lattice model introduced in the previous section, now at $\nu = 1/2$, and with additional third-neighbor density-density interactions parametrized by $V_3$, fixing $t_1 = -1$, \edit{which is related to the choice $t_1=1$ by a symmetry, as shown in the SM.}
Therefore, the parameter space is now five-dimensional (5D), $\mathbf{p}=(\lambda_{1}, t_{2}, \lambda_{2}, V_{2}, V_{3})$, with the model schematically depicted in \cref{fig:figure4}a. 
Previous studies of this model~\cite{wu_zoology_2012} have shown that it can host a MR ground state in the presence of three-body repulsive interactions, but no such phase has been found in the purely two-body regime. 
We perform a targeted search for a MR state in this system by applying target-phase optimization in a $4 \times 4$ cluster at $\nu = 1/2$. 
\edit{This choice is motivated by previous studies showing that clusters of this size can already resolve spectral signatures of the MR state~\cite{wu_zoology_2012,liu_adiabatic_2013,donna2025nonabelian}.}

Due to an even-odd effect, the ground-state degeneracy of the MR state on a torus depends on the parity of the number of particles, being six for an even number and two for an odd one~\cite{read2000paired}. 
The target ground-state manifold for this cluster then consists of a sixfold quasi-degenerate set of states, which by the generalized Pauli principle are all located at $\mathcal{K}^* = (0, 0)$, and therefore $d^* = 6$. 
We employ two subsequent refinement steps, using the target-phase loss function for a $5 \times 4$ and $6 \times 4$ cluster in order to maximize the many-body gap at large system sizes. 
The degeneracy data for such clusters is $\mathcal{K}^* = \{ (0, 0), (0, 2) \}, d^* = \{ 2, 4 \}$, and $\mathcal{K}^* = \{ (0, 0), (0, 1), (0, 2), (0, 3) \}, d^* = \{ 1, 2, 1, 2 \}$, respectively. 
We use a learning rate of $0.02$ for the optimization step.

Through this targeted search, we arrive at the parameter point $\mathbf{p} = (\lambda_{1},t_{2},\lambda_{2},V_{2}, V_{3}) = (-0.952, 0.956, -0.369, 0.781, 0.987)$, which corresponds to a $C = -1$ band hosting a non-Abelian FCI ground state. 
On a $6 \times 5$ cluster, the ED spectrum shown in \cref{fig:figure4}b reveals a twofold quasi-degenerate ground state separated from the excited manifold by a clear many-body gap, with ground-state momentum sectors matching those predicted by the generalized Pauli principle.
Adiabatic flux insertion along $\mathbf{b}_1$, shown in \cref{fig:figure4}c, reveals robust spectral flow within the quasi-degenerate subspace; the states remain separated from excited states by a large gap and return to their initial configuration only after $4\pi$ flux insertion. 
Further evidence for the MR nature of the ground state comes from the particle-cut entanglement spectrum on the $6 \times 5$ cluster (\cref{fig:figure4}d), which shows a clear entanglement gap and a number of levels below it in accordance with quasihole counting rules heuristically expected for the MR state~\cite{bernevig_emergent_2012}. 
In the SM we also show that this phase is adiabatically connected to the ground state of the Hamiltonian on the kagome lattice with repulsive three-body interactions~\cite{wu_zoology_2012}, which both further supports a non-Abelian FCI ground state and singles out the MR Pfaffian state, rather than its particle-hole conjugate~\cite{levin2007pfaffian}.
\edit{We also show in the SM the full energy spectra across several system sizes, all of which display ground-state degeneracy in the expected momentum sectors. Meanwhile, the particle entanglement gap remains large for the largest clusters we can simulate, across different system sizes and particle cuts.
Finally, we compute the many-body Chern number through a $4 \times 4$ flux grid on both the $28$- and $30$-site clusters, finding respectively $\mathcal{C} = 1/2 + 2 \times 10^{-6}$ and $\mathcal{C} = 1/2 - 9 \times 10^{-15}$, implying a Hall conductance $\sigma_{xy} = 1/2 \, e^2/h$.}
Taken together, these features offer \edit{strong numerical signatures consistent with} a fractionalized phase supporting non-Abelian anyons, in the presence of purely finite-range two-body interactions\footnote{Once again, here the refinement step leads to a final point with positive loss on the $4 \times 4$ cluster, the system size used in the main optimization step.}.

We also note that the phase shows pronounced finite-size effects, with the gap a non-monotonic function of the system size, as seen in \cref{fig:figure4}e. 
In the SM, we show that this stems from a high sensitivity to the aspect ratio, by presenting \edit{several} additional spectra using tilted clusters~\cite{repellin2014z2}, in which the gap is enhanced and the quasi-degeneracy more pronounced.
This behavior likely stems from competing states that benefit from more anisotropic cluster geometries, such as charge density waves~\cite{foglerStripeBubblePhases2001a}, as well as the fact that non-Abelian states typically have smaller energy gaps. 

\begin{figure}
    \centering
    \includegraphics[scale=1]
    {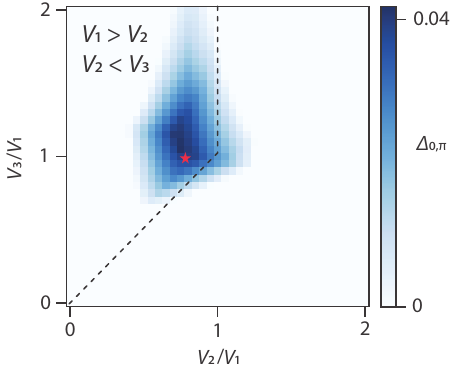}
    \caption{%
    Gap $\Delta_{0, \pi} = \min(\Delta_0, \Delta_\pi)$ as a function of $(V_2, V_3)$ at fixed $(\lambda_1, t_2, \lambda_2) =(-0.952, 0.956, -0.369)$ for a $5 \times 4$ cluster.  
    Dashed lines bound the region in which the first LL-like interaction hierarchy is realized.
    Red star indicates the parameter point hosting the non-Abelian FCI investigated here.
    }
    \label{fig:figure5}
\end{figure}

Interestingly, the parameter point found exhibits an unusual interaction hierarchy: while $V_1 > V_2$, we find $V_2 < V_3$. 
In \cref{fig:figure5}, we explore this behavior in more detail and show that the MR state seems to be most stable when these inequalities are satisfied. 
This suggests that the short-range part of the interaction must be attractive, while longer-range ones must be repulsive, a regime that can be probed in cold-atom systems~\cite{chin2010feshbach}.
\edit{The attractive character is due to $V_2<V_3$ implying that electrons lower their energy by moving from third- to second-neighbor separation, even when both are positive.}
To understand this feature, we analyze the quantum geometry of the single-particle wave functions at the optimized point, finding a trace violation $T \approx 0.21$ and Berry curvature fluctuations $\sigma_B \approx 0.32$, indicating lowest-LL-like single-particle wavefunctions. 
In the continuum, it is known that Coulomb interactions projected onto the first LL acquire precisely this kind of effective interaction hierarchy---short-range attractive and long-range repulsive---due to the nodal structure of the first LL wavefunctions~\cite{yoshioka2013quantum}.
In our lattice setting, we interpret this result as an emergent analog: since the wavefunctions are already lowest-LL-like, the bare, unprojected interactions must mimic the behavior of projected Coulomb interactions in order to stabilize the MR phase.

Finally, we note that while many properties of the phase strongly resemble those of the MR state in the first LL, there are indications that the excited-state structure is distinct. 
In particular, we observe that the low-lying excited states do not conform to known patterns of neutral excitations in the continuum MR state~\cite{reddy2024moire}.
This suggests that the energetic hierarchy of anyon excitations in this FCI may be distinct, which may have important implications for doping such a state~\cite{shi2025doping}.

\section{Discussions}
In this work, we have introduced a general, gradient‐based framework for targeting quantum phases of matter. 
By defining a target-phase loss function $\mathcal{L}$ in parameter space that directly encodes the physical signatures of the desired phase, we transform the phase‐search problem into a tractable optimization task. 
We have provided several examples of such loss functions, \edit{both in the main text and in the SM, for targeting topologically ordered phases, density waves and Fermi liquids}. 
Applying this method to spinless fermions on the kagome lattice, we have demonstrated for the first time the existence of non-ideal Abelian FCIs whose single‐particle quantum geometry challenges the expectation of LL physics mimicry, as well as non-Abelian FCIs stabilized purely by finite-range two-body interactions. \edit{Furthermore, applying the method to a generalized ten-parameter three-band model, we identify an FCI with $T \approx 12.08$ and $\sigma_B \approx 2.89$, demonstrating that FCI stability can persist under substantially stronger deviations from ideal quantum geometry than those obtained in the kagome model.}
These findings are corroborated by extensive ED calculations, including large-system spectra, adiabatic flux insertion, and entanglement spectroscopy.
While these two examples illustrate the power of the approach, the framework itself is far more general and provides a systematic route to targeting many other phases.

Looking ahead, a deeper theoretical understanding of the origin and stability of such FCIs remains a pressing challenge. It will be crucial to establish in what respects these states are distinct from their more standard counterparts, and under what circumstances they can appear. Beyond FCIs, the flexibility of our optimization method makes it applicable to a broad array of strongly correlated phases, particularly those involving fractionalization.
\edit{In that respect, developing a general recipe for constructing a target-phase loss function for a given phase of interest is an important direction for future research. Furthermore, while the present work uses phase signatures specified in advance, extending the method toward genuinely unsupervised quantum phase discovery would be an exciting direction for future work.} Finally, although we have briefly discussed extensions to auxiliary-field quantum Monte Carlo in the SM, the exponential computational cost of ED motivates combining our approach with other many-body techniques, including density matrix renormalization group~\cite{white_density_1992}, Hartree--Fock, and Monte Carlo-based methods~\cite{ceperley_qmc_1986, carleo_nnvmc_2017}.

Crucially, this work highlights the value of rethinking many-body physics through the lens of optimization and data-driven discovery. 
Our approach thus aligns naturally with developments in artificial intelligence and machine learning, suggesting a path toward increasingly autonomous and scalable exploration of quantum matter~\cite{carleo2019ML, Dawid_2025, liao2019tn, zhang2023admc, carleo_nnvmc_2017, carrasquilla2017machine}.
The integration of such frameworks with modern many-body solvers opens the door for systematic, high-throughput searches across complex Hamiltonian landscapes.

\section{Acknowledgments}
We thank Zhuo Chen, Di Luo, Andrew Ma, Junkai Dong, Tomohiro Soejima, Ahmed Abouelkomsan and Aidan Reddy for stimulating discussions.
Some numerical calculations were performed using the
\textsc{DiagHam} libraries~\cite{diagham} for which we are grateful to the contributors.
We acknowledge support from the National Science Foundation under Cooperative Agreement PHY-2019786 (The NSF AI Institute for Artificial Intelligence and Fundamental Interactions).
A.G.F.\ acknowledges support from the Whiteman Fellowship and the Surpina and Panos Eurnekian Nanotechnology Fund Fellowship. 
E.W. acknowledges funding from the MIT Undergraduate Research Opportunities Program (UROP) and the John Reed Fund.
S.V. and M.S.\ acknowledge support from the U.S.\ Office of Naval Research (ONR) Multidisciplinary University Research Initiative (MURI) under Grant No.\ N00014-20-1-2325 on Robust Photonic Materials with Higher-Order Topological Protection.
This material is based upon work also supported in part by the U.S.\ Army Research Office through the Institute for Soldier Nanotechnologies at MIT, under Collaborative Agreement Number W911NF-23-2-0121.
The MIT SuperCloud and Lincoln Laboratory Supercomputing Center provided computing resources that contributed to the results reported in this work. P.J.L.\ is supported by the MIT Pappalardo Fellowship in Physics. A.V. is supported by the
Simons Collaboration on Ultra-Quantum Matter, which is a grant from the Simons Foundation (651440, A.V.).
Any use of generative AI in this manuscript adheres to ethical guidelines for use and acknowledgement of generative AI in academic research. 
Each author has made a substantial contribution to the work, which has been thoroughly vetted for accuracy, and assumes responsibility for the integrity of their contributions~\cite{mann2024ai}.

\bibliography{references}
\end{document}


\title{\texorpdfstring{
        SUPPLEMENTAL MATERIAL\\[1ex]
        Gradient-based inverse design of quantum phases
        }
        {Supplemental Material}
       }

\author{André Grossi Fonseca}
\affiliation{Department of Physics, Massachusetts Institute of Technology, Cambridge, Massachusetts 02139, USA}
\affiliation{The NSF Institute for Artificial Intelligence and Fundamental Interactions}
\author{Eric Wang}
\affiliation{Department of Physics, Massachusetts Institute of Technology, Cambridge, Massachusetts 02139, USA}
\author{Sachin Vaidya}
\affiliation{Department of Physics, Massachusetts Institute of Technology, Cambridge, Massachusetts 02139, USA}
\affiliation{The NSF Institute for Artificial Intelligence and Fundamental Interactions}
\affiliation{Research Laboratory of Electronics, Massachusetts Institute of Technology, Cambridge, Massachusetts 02139, USA}
\author{Patrick J. Ledwith}
\affiliation{Department of Physics, Harvard University, Cambridge, Massachusetts 02138, USA}
\affiliation{Department of Physics, Massachusetts Institute of Technology, Cambridge, Massachusetts 02139, USA}
\author{Ashvin Vishwanath}
\affiliation{Department of Physics, Harvard University, Cambridge, Massachusetts 02138, USA}
\author{Marin Solja\v ci\'c}
\affiliation{Department of Physics, Massachusetts Institute of Technology, Cambridge, Massachusetts 02139, USA}
\affiliation{The NSF Institute for Artificial Intelligence and Fundamental Interactions}
\affiliation{Research Laboratory of Electronics, Massachusetts Institute of Technology, Cambridge, Massachusetts 02139, USA}

\maketitle

\setlength{\parindent}{0em}
\setlength{\parskip}{.5em}

\noindent{\small\textbf{\textsf{CONTENTS}}}\\ 
\twocolumngrid
\begingroup
    \let\bfseries\relax 
    \deactivateaddvspace 
    \deactivatetocsubsections 
    \makeatletter\@starttoc{toc}\makeatother 
\endgroup
\onecolumngrid

\count\footins = 1000 
\interfootnotelinepenalty=10000 

\section{Gradient-based search of charge-density waves}

In the main text, we focus on a target-phase loss function designed to directly identify ground-state degeneracy from the low-energy spectrum.
However, such data can only be precisely computed in ED, which is inherently limited by the exponential growth of Hilbert spaces.
Furthermore, a spectrum-based loss function may not capture the essential features needed for identifying arbitrary quantum phases.
Therefore, in this section we discuss alternative definitions of target-phase loss functions, focusing on charge-density waves (CDWs) as a case study.
We employ both ED and auxiliary-field quantum Monte Carlo to evaluate the loss in two distinct models, which in both cases are able to identify CDWs in only a few iterations.

\subsection{Target-phase loss function for charge-density waves}

A CDW ground state is defined by periodic modulations of the charge density, with modulation wavevectors $\{ \mathbf{q}_i^*\}$.
A standard diagnostic of charge order is the (connected) static structure factor, which at parameter point $\mathbf{p}$ can be written as 
\begin{equation}
    S^{\alpha\beta}_\mathbf{q}(\mathbf{p}) = \langle \, \rho^\alpha_\mathbf{q} \,\rho^\beta_{-\mathbf{q}}  \rangle - \delta_{\mathbf{q}, \boldsymbol{0}} \langle \,\rho^\alpha_\mathbf{q}  \rangle \langle \,\rho^\beta_{-\mathbf{q}}  \rangle,
\label{eq:Sq}
\end{equation}
where expectation values are taken with respect to the ground state $|\Psi_0(\mathbf{p})\rangle$, $\rho^\alpha_\mathbf{q} = \sum_{\mathbf{k}, \sigma} c^\dagger_{\mathbf{k} + \mathbf{q}, \sigma, \alpha} c^{\phantom{}}_{\mathbf{k}, \sigma, \alpha}$ is the density operator in momentum space, and $c^{\phantom{}}_{\mathbf{k}, \sigma,\alpha}\, (c_{\mathbf{k}, \sigma, \alpha}^\dag)$ is the fermion annihilation (creation) operator at momentum $\mathbf{k}$, spin $\sigma$ and pseudospin $\alpha$.
For a CDW ground state, the structure factor strongly peaks at the ordering wavevectors: $S_{\mathbf{q} \in \{ \mathbf{q}_i^*\}} \sim O(N), \, S_{\mathbf{q} \notin \{ \mathbf{q}_i^*\}} \sim O(1)$, with $N$ the system size.
Therefore, the structure factor is a good candidate for a target-phase loss function for charge order.

However, note that $S_{\mathbf{q}}$ is discontinuous at parameter points where there are level crossings in the ground state, which hinders gradient-based minimization methods.
Our solution is to employ instead a thermally averaged structure factor, defined as
\begin{equation}
    S^{\alpha \beta}_\mathbf{q}(\mathbf{p}, \beta) = \frac{1}{Z} \sum_n e^{-\beta E_n} \left( \langle \Psi_n| \, \rho^\alpha_\mathbf{q} \,\rho^\beta_{-\mathbf{q}}  |\Psi_n\rangle - \delta_{\mathbf{q}, \boldsymbol{0}} \langle \Psi_n| \,\rho^\alpha_\mathbf{q}|\Psi_n  \rangle \langle \Psi_n| \,\rho^\beta_{-\mathbf{q}}|\Psi_n  \rangle \right),
\label{eq:Sq_beta}
\end{equation}
where $|\Psi_n (\mathbf{p})\rangle$ is an eigenstate with energy $E_n(\mathbf{p})$, $Z(\mathbf{p}) = \sum_n e^{-\beta E_n(\mathbf{p})}$ is the partition function and $\beta$ the inverse temperature.
This definition reinstates continuity, since all eigenstates enter the structure factor for finite $\beta$, and directly connects to finite-temperature calculations, which can be reliably carried out with quantum Monte Carlo in sign-problem-free systems.
The last step is to sum over pseudospin indices, which can be carried out in different ways depending on the problem at hand.
Generally, we write $S_\mathbf{q}(\mathbf{p}, \beta) = \sum_{\alpha,\beta} c_{\alpha\beta}\, S^{\alpha \beta}_\mathbf{q}(\mathbf{p}, \beta)$, and elaborate below on the choice of coefficients $c_{\alpha\beta}$.

With these definitions, our goal is to define a target-phase loss function for CDWs whose global minima maximize peaks in the finite-temperature structure factor at the ordering wavevectors, and minimize the peaks at all other wavevectors. 
To do so, we take the target and complement manifolds to be $S_\mathbf{q}(\mathbf{p}, \beta)$ evaluated at, respectively, the ordering wavevectors $\{\mathbf{q}_i^*\}$ and all other momenta. 
Then, the target-phase loss function is   
\begin{equation}
\label{eq:loss_Sq}
    \mathcal{L}(\mathbf{p}, \beta) = \max_{S \in\, \operatorname{complement}}S(\mathbf{p}, \beta)  - \min_{S \in\, \operatorname{target}}S(\mathbf{p}, \beta).
\end{equation}
Note that the $\max$ and $\min$ labels are swapped with respect to the loss function for ground-state degeneracy, since here we seek to \emph{maximize} the structure factor at the target wavevectors, rather than \emph{minimize} the lowest energy levels in the target symmetry sectors, as done in the main text.

\subsection{Exact diagonalization in the kagomé lattice model}

We first apply the target-phase loss function in \cref{eq:loss_Sq} to search for CDWs using exact diagonalization. 
We employ the kagome lattice model in the main text at $\nu = 1/3$ with up to third-neighbor density-density interactions, which has a five-dimensional parameter space $\mathbf{p}=(\lambda_{1}, t_{2}, \lambda_{2}, V_{2}, V_{3})$.
At this filling, the most natural charge ordering pattern is a tripling of the unit cell, in which case the ordering wavevectors are the $K$ and $K'$ points, so $\mathbf{q}^* = \{(2\pi/3, 2\pi/\sqrt{3}), \,(4\pi/3, 0)\}$.
For our searches, we choose a $3 \times 3$ cluster, which respects the point group symmetries of the lattice and contains the ordering wavevectors.
We trace over the pseudospin indices to focus on charge order within a fixed sublattice, \ie $c_{\alpha\beta} = \delta_{\alpha\beta}$.
We fix the inverse temperature $\beta = 1$, employ the Adam optimizer with learning rate $0.25$ and compute gradients via automatic differentiation.

\begin{figure*}
    \centering
    \includegraphics[scale=1]{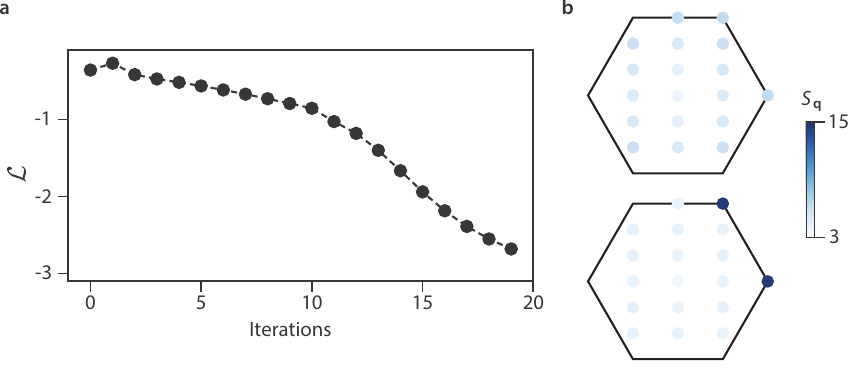}  
    \caption{%
    Optimization of CDW target-phase loss function using exact diagonalization in the kagome lattice model in the main text.
    (a) Loss function $\mathcal{L}(\mathbf{p}, \beta=1)$ as a function of number of iterations, using a $3 \times 3$ cluster. (b) Structure factor in the Brillouin zone using a $3 \times 6$ cluster for the initial (top) and final (bottom) parameter points, corresponding respectively to $(\lambda_1, t_2, \lambda_2, V_2, V_3) = (1, 0, 0, 0, 0)$ and $(\lambda_1, t_2, \lambda_2, V_2, V_3) = (1.128, -3.942, -1.277, -1.591,  4.613)$.}
    \label{fig:figure2_new}
    
\end{figure*}

Our results are shown in \cref{fig:figure2_new}.
The optimization of the target-phase loss function in \cref{eq:loss_Sq} is initialized at the point $(\lambda_1, t_2, \lambda_2, V_2, V_3) = (1, 0, 0, 0, 0)$, for which the ground state is an FCI and consequently the structure factor shows no large peaks (\cref{fig:figure2_new}b, top).
However, after less than 20 iterations the algorithm finds a point in parameter space with large negative loss (\cref{fig:figure2_new}a), located at $(\lambda_1, t_2, \lambda_2, V_2, V_3) = (1.128, -3.942, -1.277, -1.591,  4.613)$, for which the structure factor peaks at the $K$ and $K'$ points (\cref{fig:figure2_new}b, bottom), indicating a CDW with tripled unit cell.

\subsection{Auxiliary-field quantum Monte Carlo in the bilayer Hubbard model}

We now apply our method to search for CDWs using auxiliary-field quantum Monte Carlo, specifically as implemented in \textsc{alf}~\cite{alf1, alf2}.
We consider the bilayer Hubbard model for spin-1/2 fermions on the square lattice, whose Hamiltonian reads 
\begin{equation}
\begin{split}
&H = H_{\mathrm{kin}} + H_{\mathrm{int}}, \\
&H_{\mathrm{kin}} = -t_1\sum_{\langle i,j\rangle, \sigma} c_{i,\sigma, 1}^{\dagger}c^{\phantom{}}_{j, \sigma, 1} 
-t_2\sum_{\langle i,j\rangle, \sigma} c_{i,\sigma, 2}^{\dagger}c^{\phantom{}}_{j, \sigma, 2} -t_\perp\sum_{i, \sigma} (c_{i,\sigma, 1}^{\dagger}c^{\phantom{}}_{i, \sigma, 2} + \operatorname{h.c.}) , \\
&H_{\mathrm{int}} = U_{1}\sum_{i}\delta n^2_{i, 1} + U_{2}\sum_{i}\delta n^2_{i, 2},
\end{split}
\end{equation}
where $c^{\phantom{}}_{i, \sigma, l}\, (c_{i, \sigma, l}^\dag)$ is the fermion annihilation (creation) operator at site $i$, spin $\sigma$ and layer $l$, h.c. denotes the Hermitian conjugate, and $\delta n^{\phantom{}}_{i,l} = \sum_\sigma (c_{i, \sigma, l}^\dag\, c^{\phantom{}}_{i, \sigma, l} - 1/2)$ is the shifted density operator with respect to half filling.
The parameters are the nearest-neighbor intralayer hoppings $t_{1,2}$ for layers $l=1,2$, interlayer hopping $t_\perp$, and intralayer Hubbard interactions $U_{1, 2}$ for layers $l=1,2$.
Setting $t_1=1$ to fix the energy scale yields a 4D parameter space $\mathbf{p} = (t_2, t_\perp, U_1, U_2)$.
We focus on $\nu = 1/2$ with nearest-neighbor hoppings only, for which the lattice is bipartite and therefore free of the sign problem.

We search for CDWs forming a checkerboard pattern on the square lattice, which have ordering wavevector $\mathbf{q}^* = (\pi, \pi)$, in units of inverse lattice constant. 
In \textsc{alf} spin and pseudospin indices are summed over, so we have $c_{\alpha\beta}=1$.
We use a $4 \times 4$ square lattice, fix the inverse temperature $\beta = 5$ and use 20 Monte Carlo bins of 50 sweeps each for numerical efficiency.
\edit{
We employ the Adam optimizer with learning rate $0.1$. Unlike the ED implementations, the AFQMC solver is not differentiated automatically. Instead, the gradients are estimated using finite differences between independent Monte Carlo evaluations at nearby parameter points. Thus, this example demonstrates that target-phase optimization can be interfaced with a solver through numerically estimated gradients, but does not demonstrate automatic differentiation through AFQMC. Implementing automatic differentiation will be crucial to scale up target-phase optimization within AFQMC in high-dimensional parameter space, but is left for future work~\cite{zhang2023admc}.
}

\begin{figure*}
    \centering
    \includegraphics[scale=1]{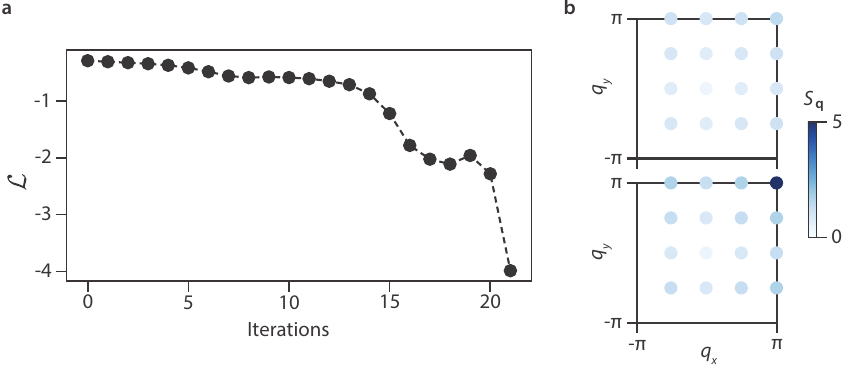}  
    \caption{%
    Optimization of CDW target-phase loss function using auxiliary-field quantum Monte Carlo in the bilayer Hubbard model.
    (a) Loss function $\mathcal{L}(\mathbf{p}, \beta=5)$ as a function of number of iterations. (b) Structure factor in the Brillouin zone for the initial (top) and final (bottom) parameter points, corresponding respectively to $(t_2, t_\perp, U_1, U_2) = (1, 1, 1, 1)$ and $(t_2, t_\perp, U_1, U_2) = (0.908, -0.540, -4.350, -4.311)$.}
    \label{fig:figure1_new}
    
\end{figure*}

Our results are shown in \cref{fig:figure1_new}.
The optimization of the target-phase loss function in \cref{eq:loss_Sq} is initialized at the point $(t_2, t_\perp, U_1, U_2) = (1, 1, 1, 1)$, for which the structure factor shows no large peaks (\cref{fig:figure1_new}b, top).
However, after only 22 iterations the algorithm finds a point in parameter space with large negative loss (\cref{fig:figure1_new}a), located at $(t_2, t_\perp, U_1, U_2) = (0.908, -0.540, -4.350, -4.311)$.
Indeed, at this point the structure factor has a large peak at $(\pi, \pi)$ (\cref{fig:figure1_new}b, bottom), indicating a checkerboard CDW.
We note that this parameter point is adiabatically connected to the region with $t_1 = t_2, t_\perp = 0, U_1 = U_2 < 0$, which corresponds to two identical decoupled Hubbard layers with attractive interactions, each of which is known to harbor a charge-ordered ground state~\cite{hirsch1985hubbard}.
\edit{
The independently established nature of this phase as a CDW provides a direct validation that minimizing the target-phase loss leads to the targeted phase, in the context of a many-body solver other than exact diagonalization.
}

\section{Gradient-based search of Fermi liquids}

\edit{We now demonstrate that target-phase loss functions can also be constructed for gapless phases by considering a Fermi liquid as an example. A defining signature of a Fermi liquid is a sharp Fermi surface, across which the zero-temperature momentum occupation $n_{\mathbf{k}}$ is discontinuous in the thermodynamic limit. On a finite cluster, this behavior appears as a strong contrast between occupied and unoccupied momenta. This motivates a loss function that directly maximizes the spread of the momentum occupation, without specifying the shape or location of the Fermi surface in advance.}

\edit{
As we did for the CDW loss function, we use the Boltzmann-weighted momentum occupation to avoid abrupt changes of the ground state at level crossings: 
\begin{equation}
    n_{\mathbf{k}}(\mathbf{p},\beta)=\frac{1}{Z}\sum_m e^{-\beta E_m(\mathbf{p})}\langle\Psi_m(\mathbf{p})|\,c_{\mathbf{k}}^\dagger c_{\mathbf{k}}^{\phantom{\dagger}}\,|\Psi_m(\mathbf{p})\rangle,
    \label{eq:nk_beta}
\end{equation}
where $c_{\mathbf{k}}^{\phantom{\dagger}}$ annihilates a fermion in the lowest band at momentum $\mathbf{k}$, $|\Psi_m(\mathbf{p})\rangle$ is a many-body eigenstate with energy $E_m(\mathbf{p})$, $Z(\mathbf{p})=\sum_m e^{-\beta E_m(\mathbf{p})}$ is the partition function, and $\beta$ is the inverse temperature. The target-phase loss function for Fermi liquids is then
\begin{equation}
    \mathcal{L}(\mathbf{p},\beta)=\min_{\mathbf{k}} n_{\mathbf{k}}(\mathbf{p},\beta)-\max_{\mathbf{k}} n_{\mathbf{k}}(\mathbf{p},\beta).
    \label{eq:loss_nk}
\end{equation}
Because $0\leq n_{\mathbf{k}}\leq1$, this loss is bounded as $-1\leq \mathcal{L}\leq 0$. 
Minimizing \cref{eq:loss_nk} increases the contrast between the most and least occupied momenta and therefore is a good proxy for the development of a sharp Fermi surface.} 

\edit{We apply this loss function using exact diagonalization in the kagome lattice model of the main text. Because we now retain the full single-particle dispersion of the lowest band, $V_1$ does not simply rescale the spectrum; therefore, we keep it as a tunable parameter and set $t_1 = 1$. We consider a $4\times4$ cluster at filling $\nu=1/4$ and set $\beta=10$. We optimize the six-dimensional parameter vector $\mathbf{p}=(\lambda_1,t_2,\lambda_2,V_1,V_2,V_3)$ using Adam with learning rate $0.001$, with gradients computed by automatic differentiation.}

\begin{figure}
    \centering
    \includegraphics[scale=1]{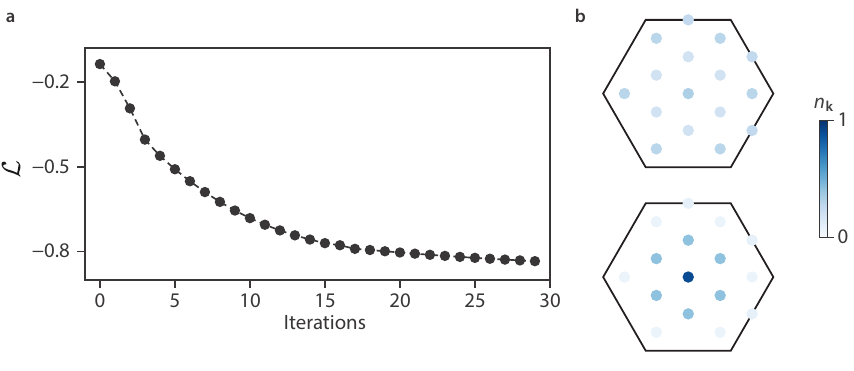}
    \caption{
    \edit{Optimization of the Fermi-liquid target-phase loss function using exact diagonalization in the kagome lattice model. (a) Loss function $\mathcal{L}(\mathbf{p},\beta=10)$ as a function of the number of Adam iterations for a $4\times4$ cluster at filling $\nu=1/4$. (b) Boltzmann-weighted momentum occupation $n_{\mathbf{k}}(\beta=10)$ in the Brillouin zone at the initial (top) and final (bottom) parameter points, corresponding respectively to $\mathbf{p}=(0.3,-0.3,0.2,1,0,0)$ and $\mathbf{p}=(0.282,-0.280,0.182,0.974,-0.018,-0.016)$.}
    }
    \label{fig:figure2_nk}
\end{figure}

\edit{The results are shown in \cref{fig:figure2_nk}. The optimization is initialized at $\mathbf{p}=(0.3,-0.3,0.2,1,0,0)$, at which the lower band is very flat~\cite{tang_high-temperature_2011}, thereby disfavoring a Fermi liquid; there, the momentum occupation is fairly uniform and the loss is close to zero [\cref{fig:figure2_nk}(b), top]. Over 30 iterations, the loss decreases rapidly at first and then more gradually, reaching values near the minimum value of $-1$ [\cref{fig:figure2_nk}(a)] at parameter point $\mathbf{p}=(0.282,-0.280,0.182,0.974,-0.018,-0.016)$. At this point, the momentum occupation develops a pronounced separation between high- and low-occupation regions in the Brillouin zone [\cref{fig:figure2_nk}(b), bottom], consistent with a sharp Fermi surface.
We have also checked that the energy spectrum has no clear many-body gap, indicating a gapless phase.}

\section{Band structure for non-ideal FCI}

\begin{figure*}
    \centering
    \includegraphics[scale=1]
    {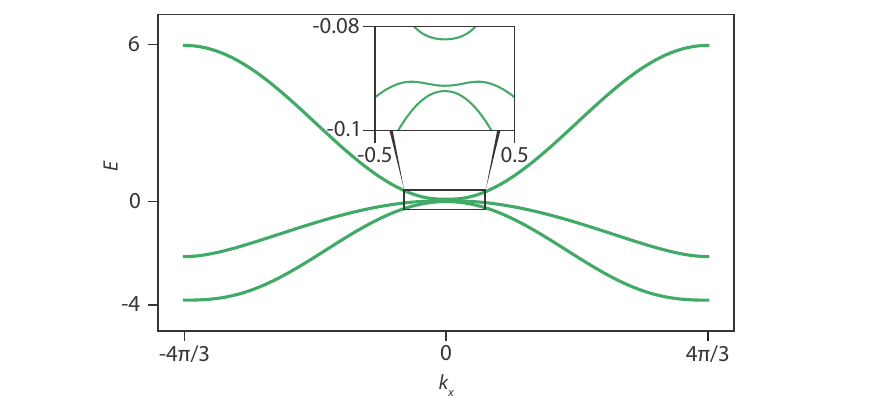}
    \caption{%
    Single-particle band structure at $k_y = 0$ for the kagome model at parameters $(t_1, \lambda_1, t_2, \lambda_2) = (1, -0.169, -0.992, -0.157)$.
    Inset shows the gap between bands and quadratic nature of the near band touching.
    }
    \label{fig:figure1}
\end{figure*}

In \cref{fig:figure1} we show the band structure for the kagome model considered in the main text at the parameter point hosting a non-ideal FCI.
We have checked that the two lowest bands are gapped everywhere, with the small gap at the $\Gamma$ point being responsible for the concentrated Berry curvature $\mathcal{F}(\mathbf{k})$ and momentum-resolved trace violation $T(\mathbf{k})$.
\edit{Furthermore, because of this concentrated band geometry, calculations of the trace violation and Berry curvature fluctuations require very dense momentum meshes; here, we used a $1000 \times 1000$ momentum grid, and have explicitly checked that this is sufficient for convergence.}

However, as seen in the main text, $\mathcal{F}(\mathbf{k})$ vanishes at $\Gamma$ and peaks at a ring around it. 
This is consistent with a slightly gapped quadratic degeneracy, as shown in~\cref{fig:figure1}.
To see this, we consider a $\mathbf{k} \cdot \mathbf{p}$ expansion of a generic 2D two-band Bloch Hamiltonian about a near-quadratic band touching:
\begin{equation}
\label{eq:kdotp}
    h(\mathbf{k}) = \begin{pmatrix}
    m & \alpha\,(k_x -i k_y)^2\\
    \alpha\,(k_x +i k_y)^2 & -m
    \end{pmatrix}
\end{equation}
where $\alpha$ and $m$ are parameters from the Bloch Hamiltonian.
Rewriting in terms of Pauli matrices
\begin{equation}
\label{eq:kdotp_pauli}
     h(\mathbf{k}) = d_x \sigma_x + d_y \sigma_y + d_z \sigma_z, \quad d_x = \alpha\,(k_x^2 - k_y^2), \, d_y = 2\alpha k_x k_y, \, d_z = m
\end{equation}
The Berry curvature for the lower band can be written solely in terms of the components of the $\mathbf{d}$ vector as
\begin{equation}
\label{eq:berry_curv_general}
    \mathcal{F}(\mathbf{k}) = \frac{1}{2d^3} \sum_{\{\{ijk\}\}} d_i (\partial_{x}d_j\,\partial_y d_k - \partial_{y}d_j\,\partial_x d_k),
\end{equation}
where $d = \sqrt{d_x^2 + d_y^2 + d_z^2}$, $\partial_i \equiv \partial_{k_i}$ and the sum is performed over cyclic permutations of $\{ijk\}$.
Plugging \cref{eq:kdotp_pauli} into \cref{eq:berry_curv_general} yields
\begin{equation}
\label{eq:berry_curv}
    \mathcal{F}(\mathbf{k}) = \frac{2m\alpha^2 k^2}{(m^2 + \alpha^2 k^4)^{3/2}},
\end{equation}
which vanishes at the origin and peaks at a ring near $k = \sqrt{m/\alpha}$, as claimed.

\section{Non-ideal FCI momentum-space occupation}

\begin{figure}
    \centering
    \includegraphics[scale=1]
    {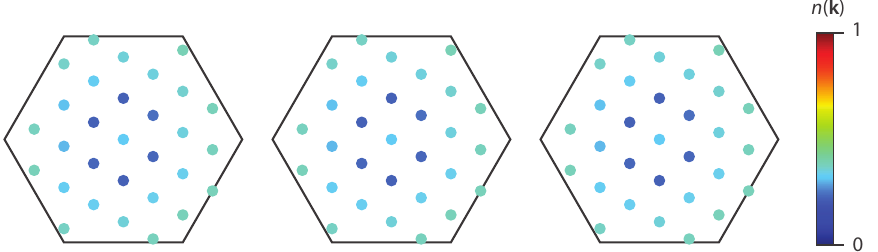}
    \caption{%
    Occupation numbers $n(\mathbf{k})$ for the three quasi-degenerate ground states of the non-ideal FCI, calculated from a $6 \times 5$ cluster.
    }
    \label{fig:figure2}
\end{figure}

In \cref{fig:figure2} we show the occupation numbers $n(\mathbf{k}) = \bra{\Psi_i}c^\dag_\mathbf{k} c_\mathbf{k} \ket{\Psi_i}$ for each of the three quasi-degenerate ground states $\Psi_i$ at the parameter point hosting the non-ideal FCI.
The momentum-space distribution of $n(\mathbf{k})$ is largely homogeneous and close to $1/3$ for all points, which provides further evidence for the FCI ground state.
However, we note that, in general, an analysis of the local order parameter $n(\mathbf{k})$ cannot by itself single out a topological phase.

\section{Extended numerical data for competing phases}

\begin{figure}
    \centering
    \includegraphics[scale=1]
    {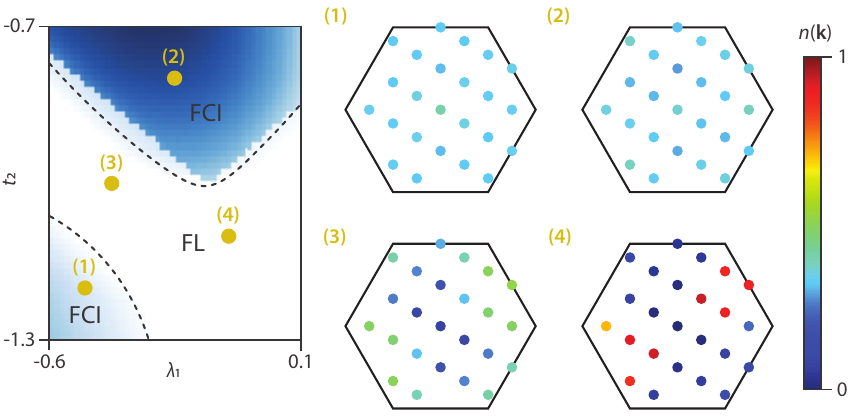}
    \caption{%
    Occupation numbers $n(\mathbf{k})$ for selected points in the phase diagram for the plane $(\lambda_2, V_2) = (-0.157, 0.25)$.
    Points (1), (2), (3), (4) are located at parameters $(\lambda_1, t_2) = (-0.5, -1.2),\, (-0.25, -0.8),\, (-0.425, -1.0),\, (-0.1, -1.1)$, respectively.
    }
    \label{fig:figure3}
\end{figure}

In \cref{fig:figure3} we show $n(\mathbf{k})$ for a few selected points in the phase diagram in the main text.
While $n(\mathbf{k})$ is largely uniform within the FCI regions, it displays a sharp discontinuity in a region near $\Gamma$ for the non-FCI regions, signaling the presence of a Fermi surface consistent with a Fermi liquid phase.
We have also checked that the spectra are gapless at these points under flux insertion.
Within the parameter ranges shown, we have not found evidence of any other competing phases, such as charge density waves.

\section{Tuning between non-ideal and ideal bands}

\begin{figure}
    \centering
    \includegraphics[scale=1]
    {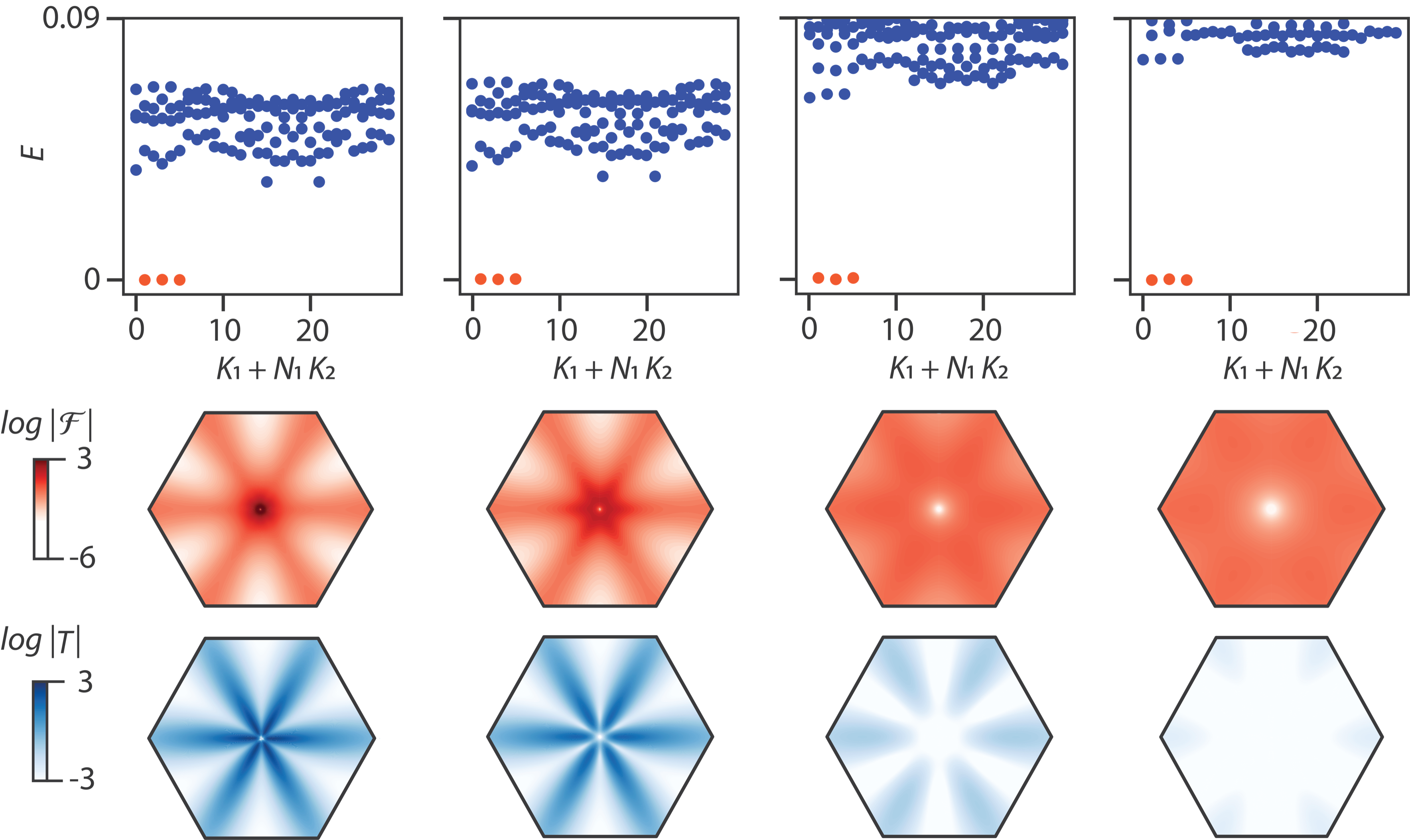}
    \caption{%
    ED spectrum for a $6 \times 5$ cluster as a function of center-of-mass momentum (top), and single-particle Berry curvature $\mathcal{F}(\mathbf{k})$ (middle) and momentum-resolved trace violation $T(\mathbf{k})$ (bottom) distributions, as both $T$ and $\sigma_B$ are decreased towards ideal values.
    We only plot the lowest four energy levels per momentum sector.
    The logarithm of $\mathcal{F}(\mathbf{k})$ and $T(\mathbf{k})$ is shown for visual clarity.
    Values of $(T, \sigma_B)$ are (from left to right): $(3.3, 5.2),\, (2.6, 2.1),\, (0.7, 0.6),\, (0.3, 0.3)$. 
    Parameters: $t_1 = V_1 = 1, V_2 = 0.25$ and (from left to right): $(\lambda_{1}, t_{2}, \lambda_{2}) = (-0.169, -0.992, -0.157),\, (-0.166, -0.987, -0.159),\, (-0.217, -0.865, -0.210),\, (-0.275, -0.726, -0.269)$.
    }
    \label{fig:figure4}
\end{figure}

In \cref{fig:figure4} we show ED spectra for some selected points along the path from the non-ideal point to the ideal region, obtained with gradient descent on $T$.
Both $\mathcal{F}(\mathbf{k})$ and $T(\mathbf{k})$ become smoother, and $\sigma_B$ and $T$ decrease until they are near zero. 
The many-body gap remains open throughout the path, which shows that this non-ideal FCI point is adiabatically connected to ideal FCIs.

\section{Non-ideal FCIs in general three-band models}

In this section, we generalize the kagome model in the main text in order to test the target-phase loss function method in a high-dimensional parameter space.

\subsection{Model}

\begin{figure}
    \centering
    \includegraphics[scale=1]{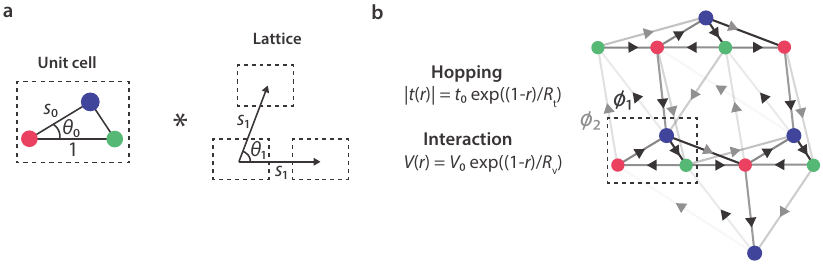}
    \caption{%
    The general class of three-band models with ten parameters.
    (a) Geometric parameters defining the unit cell and lattice.
    (b) Exponentially decaying hoppings and density-density interactions, with bond color denoting strength.
    Hopping phases $\phi_1, \phi_2$ are also included in a similar way to those of the kagome model.}
    \label{fig:figure_6_new}
\end{figure}

We first allow geometric deformations to the lattice, as shown in \cref{fig:figure_6_new}a. We consider models with three sublattices A, B and C, with the intra-unit-cell bond distances between the AB and AC sublattices set to unity and $s_0$, respectively, and with bond angle $\theta_0$ between them.
The size of the primitive lattice vectors is $s_1$ and their subtended Bravais angle is $\theta_1$.
Note that, apart from lattices with primitive lattice vectors of unequal size, these variables parametrize \emph{every} possible 2D lattice with three sublattices.
We choose lattice vectors of equal magnitude to avoid strong finite-size effects in ED that may stem from highly inhomogeneous lattices.

\begin{figure*}
    \centering
    \includegraphics[scale=1]{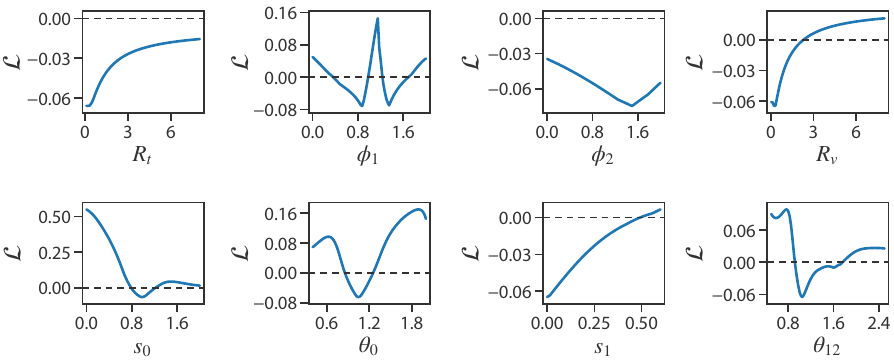}  
    \caption{%
    \edit{
    One-dimensional sweeps of the target-phase loss $\mathcal{L}$ around an FCI for the generalized three-band model, computed at system size $3 \times 3$. Dashed horizontal lines denote $\mathcal{L}=0$ and therefore bound the FCI regions at this system size.
    Throughout we fix $t_0 = -1, V_0 = 1$.
    }
    }
    \label{fig:loss_sweeps}
\end{figure*}

We incorporate these geometric variables into the tight-binding Hamiltonian by considering distance-based hoppings and density-density interactions, of the form $t(\mathbf{r}) = t_0 \exp((1-r)/R_t), \, V(\mathbf{r}) = V_0 \exp((1-r)/R_v)$, respectively.
Because we would like to recover the kagome model when constraining this class of models, we only allow hoppings and interactions between sites that were connected in the kagome model, as illustrated in \cref{fig:figure_6_new}b.
Finally, we take the hopping phases to be arranged in precisely the same way as they were in the kagome model, which are parametrized by variables $\phi_1, \phi_2$.
Therefore, the parameter space is ten-dimensional (10D), $\mathbf{p}=(t_0, R_t, \phi_1, \phi_2, V_0, R_v, s_0, \theta_0, s_1, \theta_1)$.
In particular, we recover the 4D parameter space of the kagome model $\mathbf{p}=(\lambda_{1},t_{2},\lambda_{2},V_{2})$ if we take $t_0 = -z_1, R_t = (\sqrt{3}-1)/\ln{(z_1/z_2)}, \phi_1 = \arg{z_1}, \phi_2 = \arg{z_2} + \pi (1-\sgn (t_2))/2, V_0 = 1, R_v = (\sqrt{3}-1)/\ln{(1/V_2)}, s_0 = 1, \theta_0 = \pi/3, s_1 = 2, \theta_1 = \pi/3$, where $z_i \equiv \sqrt{t^2_i + \lambda^2_i}$.
Therefore, the parameter space for the kagome model is indeed a low-dimensional subspace of this class of three-band models.
\edit{
To illustrate the structure of the loss landscape, in \cref{fig:loss_sweeps} we show one-dimensional sweeps of the target-phase loss near an FCI. The loss is smooth over extended regions of these sweeps, but can also exhibit sharp features due to the max/min operations entering the loss function.
}

Note that not all of the 10D space is physically accessible, as we must have $R_t, R_v, s_0 >0$ and the length of the primitive lattice vectors to be larger than the intra-unit-cell vectors, \ie $s_1 > \max (1, s_0)$.
To perform our searches under these constraints, we define auxiliary variables $R_{t} = |R_{tt}|, R_{v} = |R_{vv}|, s_0 = |S_0|, s_1 = 1 + s_0 + |S_1|$, and then carry out unconstrained searches in the new space $\mathbf{p}=(t_0, R_{tt}, \phi_1, \phi_2, V_0, R_{vv}, S_0, \theta_0, S_1, \theta_1)$.

\subsection{Optimization and results}

We perform a similar optimization procedure to the one described in the main text, using a $3 \times 3$ cluster at filling $\nu = 1/3$.
However, here we divide the search into three steps: first, we optimize the target-phase loss function $\mathcal{L}(\mathbf{p})$ defined in Eq. (2) of the main text, until it attains a negative value.
Then, we find it beneficial to define a second target-phase loss function $\mathcal{L}'(\mathbf{p}) = \mathcal{L}(\mathbf{p})/\delta(\mathbf{p})$, where $\delta(\mathbf{p}) = \max_{\{ \Phi_j \}} d(\mathbf{p}, \Phi_j)$ is the maximum over the spread of the ground states, defined as $d(\mathbf{p}, \Phi) = \max_{E \in\, \operatorname{target}}E(\mathbf{p}, \Phi)  - \min_{E \in\, \operatorname{target}}E(\mathbf{p}, \Phi)$.
This loss is optimized until $\mathcal{L}'(\mathbf{p}) < -1$,
\ie the spread is smaller than the many-body gap, ensuring that the ground-state quasi-degeneracy is of high quality.
Finally, we maximize the trace violation as before, by defining $\mathcal{L}''(\mathbf{p})=-T(\mathbf{p})$.
We initialize points randomly within the unit hypercube $[0, 1]^{10}$ and employ the Adam optimizer as before, testing different learning rates $0.05$ and $0.1$ for the first two steps and fixing it to be $0.025$ for the last step.

\edit{ 
For the ED losses used during optimization, we explicitly construct the many-body Hamiltonian in each total-momentum sector and obtain the required low-energy eigenvalues using a dense Hermitian eigensolver.
At points in parameter space with non-degenerate low-energy spectrum, automatic differentiation follows the active eigenvalue and max/min branches; for the flux maximum, only the active flux contributes. At points with degeneracies, the gradients are shared equally between the two branches. However, such exact degeneracies are not generic and do not affect optimization in practice.
The trace violation $T$ is differentiated using standard gauge-invariant projectors, which are constructed directly from the eigenvalues of the Bloch Hamiltonian.
The quantum geometric tensor is then evaluated from gauge-invariant combinations of these projectors. This avoids differentiating the arbitrary phases of complex Bloch eigenvectors, while allowing automatic differentiation to propagate through the eigenvalues, projectors, and Hamiltonian matrix elements to the hopping parameters.
We have also benchmarked the gradients obtained through automatic differentiation against finite-difference derivatives and find agreement within the expected numerical accuracy.
}

\edit{
From preliminary runs, we find that this model can host highly non-ideal FCIs, often with trace violation $T > 12$.
To assess the robustness and efficiency of this procedure, we compare our gradient-based optimization method with uniform random sampling for locating an FCI with $T > 12$\footnote{We have found that simple Cartesian grid search cannot locate non-ideal FCIs, even for relatively dense grids.}.
For this benchmark, we fix $V_0=1$ to set the energy scale, leaving nine free parameters, and use the same bounded parameter space for both methods.
We sample $t_0\in[0.01,1]$, $R_t,R_v\in[0.05,1]$, $\phi_1,\phi_2,s_0,\theta_0,S_1\in[0,1]$, and $\theta_1\in[0.025,1]$, with $s_1=1+s_0+S_1$.
For our comparison purposes, a point is counted as a successful hit if it has both a negative loss on the $3 \times 3$ cluster and $T>12$.
In the next section, we characterize in more detail the thermodynamic stability of one such non-ideal FCI.
}
\par\begin{table}[ht]
    \centering
    \edit{
    \small
    \setlength{\tabcolsep}{2pt}
    \begin{tabular}{cccccccc}
        \hline\hline
        \shortstack{Learning rate} & \shortstack{FCI with\\$T>12$} & \shortstack{FCI with\\$T<12$} & \shortstack{Local\\minimum} & \shortstack{Iteration budget\\exhausted} & \shortstack{Total} & \shortstack{Wall time/\\ iteration} & \shortstack{Peak\\memory} \\
        \hline
        $0.1$ & 4 & 2 & 19 & 5 & 30 & $51\,$ms & $423\,$MiB \\
        $0.05$ & 6 & 1 & 17 & 6 & 30 & $55\,$ms & $450\,$MiB \\
        \hline\hline
    \end{tabular}
    }
    \caption{
    \edit{Outcomes of the 30 gradient-based optimization runs at each learning rate. Both benchmarks use the same initial points, and the learning rate for the final maximization of $T$ is fixed at $0.025$.}
    }
    \label{tab:gradient_outcomes}
\end{table}
The results from the gradient benchmarks are summarized in \cref{tab:gradient_outcomes}. For each learning rate, we perform 30 independent runs from the same initial points sampled uniformly from the parameter box, with optimizations carried out following the three-stage objective detailed above. Each trajectory is allowed at most 30 iterations, with all parameters projected back into the search domain after every update. A trajectory is terminated early upon reaching the target or when it becomes trapped in a local minimum, defined by a loss change of less than $10^{-4}$ between consecutive iterations within the same optimization stage. At learning rates $0.1$ and $0.05$, the method locates four and six FCIs with $T>12$, respectively, with each requiring an average of 132 and 96 loss evaluations to locate a non-ideal FCI. Thus, decreasing the learning rate to $0.05$ increases the number of hits, while slightly decreasing the number of runs stuck at local minima, which remains the main source of algorithmic failure in both cases.	

\edit{We also tracked the wall-clock times and peak memory usage for these benchmarks, which were run on a local machine using CPUs only. Averaged over all optimization iterations, the wall time per iteration is $51\,$ms for learning rate $0.1$ and $55\,$ms for learning rate $0.05$. The corresponding average peak resident-set sizes, obtained by averaging the peak process RSS measured for each of the 30 initializations, are $423\,$MiB and $450\,$MiB, respectively. This small memory difference is due to the learning rate $0.05$ batch being run second in the same notebook kernel.
These modest resource requirements are specifically due to our use of the $3 \times 3$ cluster and are expected to increase substantially for larger clusters.}

The comparative results from target-phase optimization and uniform random sampling are summarized in \cref{tab:search_benchmark}. For the random-search baseline, we evaluate 10,000 independent points drawn uniformly from the same parameter box. Of these, only six are FCIs with $T>12$, giving a hit rate of $0.06\%$ and requiring an average of 1666.7 loss evaluations per successful point. At learning rate $0.05$, gradient-based optimization requires only 96 loss evaluations per hit, approximately $17.4$ times fewer than uniform random sampling and corresponding to a $94.2\%$ reduction in the number of many-body calculations per hit. These results demonstrate a clear advantage for gradient-based optimization in this high-dimensional search space, which is expected to become increasingly consequential as the number of tunable Hamiltonian parameters grows.

\begin{table}[t]
    \caption{
    \edit{Statistical comparison between gradient-based optimization with two learning rates and uniform random sampling for locating an FCI with $T>12$. All methods use the same parameter domain and success criterion.}
    }
    \label{tab:search_benchmark}
    \edit{
    \small
    \setlength{\tabcolsep}{2pt}\begin{tabular}{lrrrrrr}
        \hline\hline
        Method & \shortstack{Points} & Hits & Hit rate & \shortstack{Loss\\evaluations} & \shortstack{Loss evaluations/\\hit} \\
        \hline
        Gradient (lr $=0.1$) & 30 & 4 & $13.33\%$ & 528 & 132.0 \\
        Gradient (lr $=0.05$) & 30 & 6 & $20\%$ & 576 & 96.0 \\
        Uniform random sampling & 10,000 & 6 & $0.06\%$ & 10,000 & 1666.7 \\
        \hline\hline
    \end{tabular}
    }
\end{table}

\subsection{Representative run}
\begin{figure}
    \centering
    \includegraphics[scale=1]
    {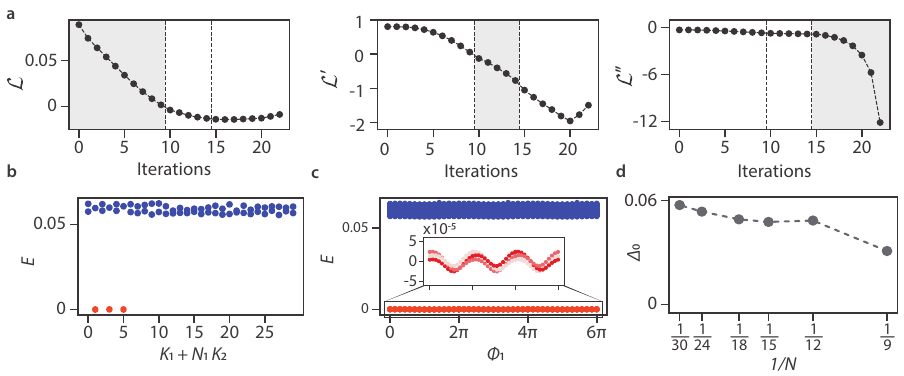}
    \caption{%
    (a) Loss functions $\mathcal{L}, \mathcal{L}', \mathcal{L}''$ as a function of number of iterations, using a $3 \times 3$ cluster.
    Shaded regions denote the specific iterations where each loss function was employed.
    The initial and final points used were respectively $(t_0, R_t, \phi_1, \phi_2, R_v, s_0, \theta_0, s_1, \theta_1) = (0.347, 0.756, 0.974, 0.460, 0.768, 0.964, 0.330, 2.520, 0.295)$ and $(0.285, 1.483, 0.949, 1.128, 0.409, 1.898, 0.330, 2.942, 0.036)$.
    (b) Zero-flux ED spectrum for a $6 \times 5$ cluster as a function of center-of-mass momentum.
    (c) ED spectrum for a $6 \times 5$ cluster as a function of magnetic flux $\Phi_1$ along reciprocal lattice vector $\mathbf{b}_1$.
    Inset: zoomed in spectrum showing spectral flow of the ground state manifold.
    (d) Zero-flux gap $\Delta_0$ as a function of inverse system size $1/N$.
    All spectral quantities are expressed in units of $V_0$.}
    \label{fig:figure_7_new}
\end{figure}

The results for a representative optimization run are shown in \cref{fig:figure_7_new}.
In \cref{fig:figure_7_new}a we plot the three loss functions along the optimization path, showing that the target-phase loss function method can efficiently locate a non-ideal FCI with only 20 iterations, converging onto a point with $T > 3$.
Remarkably, we find that a few more iterations lead to FCIs \edit{which exhibit much larger deviations from ideal quantum geometry}, reaching $T \approx 12.08$ and $\sigma_B \approx 2.89$ after a total of 22 iterations.
We confirm that the ground state is an FCI through its energy spectrum and flux insertion, both on a $6 \times 5$ cluster, as displayed in \cref{fig:figure_7_new}b,c.
Lastly, we show in \cref{fig:figure_7_new}d that the many-body gap remains finite for many system sizes, providing evidence that it is the true ground state in the thermodynamic limit.

\begin{figure}
    \centering
    \includegraphics[scale=1]
    {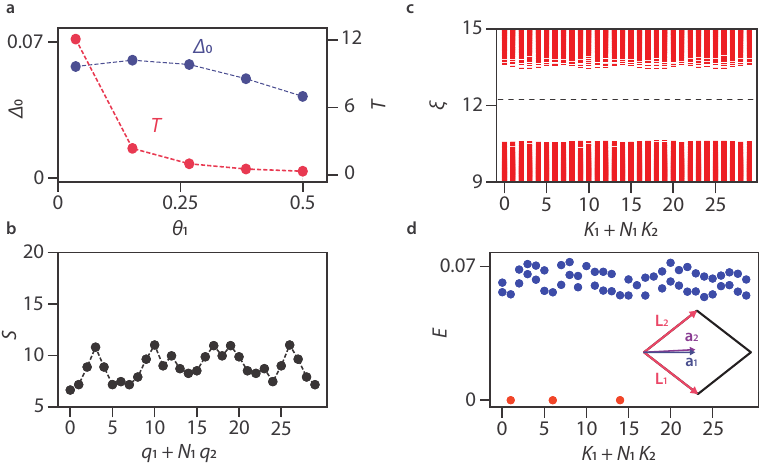}
    \caption{%
    Further evidence for a non-ideal FCI at $(t_0, R_t, \phi_1, \phi_2, V_0, R_v, s_0, \theta_0, s_1, \theta_1) = (0.285, 1.483, 0.949, 1.128, 0.276, 0.409, 1.898, 0.330, 2.942, 0.036)$.
    (a) Zero-flux gap $\Delta_0$ on a $6 \times 5$ cluster and trace violation $T$ as a function of Bravais angle $\theta_1$.
    (b) Ground-state structure factor $S_\mathbf{q}$ on a $6 \times 5$ cluster as a function of ordering momentum $\mathbf{q}$.
    (c) Particle-cut entanglement spectrum on a $6 \times 5$ cluster for subspace particle number $N_A = 4$.
    The number of states below the dashed line is 9975, consistent with a Laughlin ground state.
    (d) ED spectrum for a tilted $6 \times 5$ cluster as a function of center-of-mass momentum.
    The cluster is defined by cluster vectors $\mathbf{L}_{1,2} = n_{x1,2} \mathbf{a}_1 + n_{y1,2} \mathbf{a}_2$, with cluster integers $(n_{x1}, n_{y1}, n_{x2}, n_{y2}) = (16, -15, -14, 15)$.
    Inset: Real-space cluster used, with primitive lattice vectors $\mathbf{a}_1 = s_1 (1, 0), \mathbf{a}_2 = s_1(\cos \theta_1, \sin \theta_1)$.
    All spectral quantities are expressed in units of $V_0$.}
    \label{fig:figure_8_new}
\end{figure}

In order to investigate the origin of such a high trace violation, we first observe that the Bravais angle is quite small, $\theta_1 = 0.036$, which leads to a highly anisotropic, quasi-1D lattice.
In \cref{fig:figure_8_new}a we show the behavior of the many-body gap and trace violation as $\theta_1$ is increased, and we observe that the gap stays open throughout the process, while $T$ sharply decreases, confirming both that the shallow angle is the origin of the strong non-ideality and that the ground state is adiabatically connected to a genuinely 2D Laughlin state.
However, this adiabatic connectivity is also consistent with a quasi-1D CDW~\cite{taothouless, seidel2005}.
To rule this out, in \cref{fig:figure_8_new}b we show the ground-state structure factor $S_\mathbf{q}$, which does not exhibit any prominent peaks. We have checked that $S_\mathbf{q}$ is largely insensitive to increases of $\theta_1$ towards the 2D limit.
Likewise, the particle-cut entanglement spectrum (\cref{fig:figure_8_new}c) shows a large entanglement gap and number of levels below it matching the Laughlin count. 
Finally, in \cref{fig:figure_8_new}d we construct a highly isotropic cluster using tilted boundary conditions (see discussion in Section S8), for which we still find the expected ground-state quasi-degeneracy for an FCI.
We have checked that all of these results remain consistent for several different cluster sizes and geometries.
Taken together, these features point to an FCI \edit{with highly non-ideal quantum geometry}.
It would be interesting for future work to explore the origin of the stability of such a non-ideal FCI, possibly through recently introduced coupled-wire constructions~\cite{shavit2024}.

\section{Interpolation between two-body and three-body Hamiltonians}

\begin{figure}
    \centering
    \includegraphics[scale=1]
    {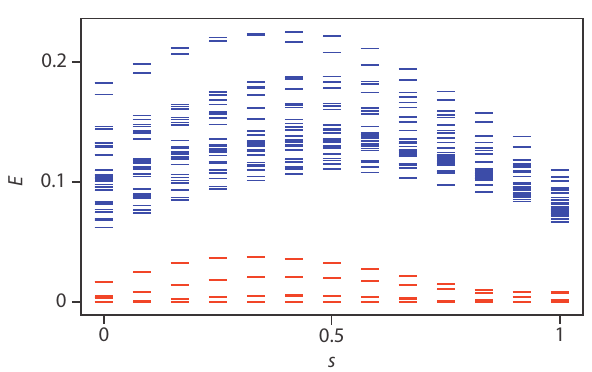}
    \caption{%
    ED spectrum for a $5 \times 4$ cluster of the interpolation Hamiltonian $H(s)$ defined in \cref{eq:interp_ham}, as a function of adiabatic parameter $s$.
    }
    \label{fig:figure5}
\end{figure}

In \cref{fig:figure5} we show an adiabatic interpolation between the Hamiltonian hosting the non-Abelian FCI with only two-body interactions, and the Hamiltonian with a purely three-body repulsive interaction.
Specifically, we consider the following interpolation Hamiltonian:
\begin{equation}
\begin{split}
\label{eq:interp_ham}
&H(s) = H_{\mathrm{kin}}(s) + H_{\mathrm{int}}(s), \\
&H_{\mathrm{kin}}(s) = -\sum_{\langle i,j\rangle} (t_{1}+i\nu_{ij}\lambda_1(s)) c_{i}^{\dagger}c_{j} 
- \sum_{\langle\langle i,j\rangle\rangle}(t_{2}(s)+i\nu_{ij}\lambda_2(s))c_{i}^{\dagger}c_{j}, \\
&H_{\mathrm{int}}(s) = V_{1}\sum_{\langle i,j\rangle}:\!n_{i}\,n_{j}\!: + V_{2}(s)\sum_{\langle\langle i,j\rangle\rangle}:\!n_{i}\,n_{j}\!: + V_{3}(s)\sum_{\langle\langle\langle i,j\rangle\rangle\rangle}:\!n_{i}\,n_{j}\!: + W(s)\sum_{\langle i,j,k\rangle}:\!n_{i}\,n_{j}\,n_{k}\!:,
\end{split}
\end{equation}
where $s$ is the interpolation parameter and $W$ is the three-body interaction strength, defined over triangles of the kagome lattice in the same way as~\cite{wu_zoology_2012}.
Note that we fix $t_1 = 1$ and $V_1 = 1$ throughout the interpolation process.
Because the non-Abelian FCI in the main text was found in the sector $t_1 = -1$, we map the parameter points to the $t_1 = 1$ sector using the following symmetries of the single-particle kagome Hamiltonian:
\begin{equation}
z_1 \rightarrow \omega z^*_1, \, z_2 \rightarrow \omega^2 z^*_2, \quad z_i \equiv t_i + i \lambda_i, \quad \omega \equiv e^{-i 2\pi/3}, 
\end{equation}
which then takes the original parameter point to $\mathbf{p}^0 = (t_1, \lambda^0_{1},t^0_{2},\lambda^0_{2}) = (1, 0.294, -0.602, 0.486)$ without modifying the spectrum\footnote{We note that this symmetry can be employed in the optimization scheme described in the main text to reduce the size of the search space.}. 
This transformation is equivalent to a unitary transformation of the operators 
\begin{equation}
    \boldsymbol{c}^{(\dag)}_i \mapsto U^{(\dag)} \boldsymbol{c}^{(\dag)}_i, \quad U = \operatorname{diag} (1, \omega, \omega^2),
\end{equation}
where $\boldsymbol{c}^{(\dag)}_i  = (c_{Ai}^{(\dag)}, c_{Bi}^{(\dag)}, c_{Ci}^{(\dag)})$ are the creation/annihilation operators for the three sublattices in the $i$-th unit cell. Additionally, the chirality of the hoppings is reversed to bring $\lambda_1$ closer to its final value.  

For the interpolation we use linear functions $\lambda_1(s) = (1-s) \lambda^0_1 + s\, \lambda^f_1, \, t_2(s) = (1-s) t^0_2 + s\, t^f_2$, \etc, where $\mathbf{p}^f = (\lambda^f_{1},t^f_{2},\lambda^f_{2},V^f_{2}, V^f_{3}) = (1, 0, 0, 0, 0)$.
For the three-body strength, we take $W^0 = 0, \, W^f = 200$ so that the many-body gaps are roughly of the same size.
Because the gap stays open throughout the interpolation, the ground state of the two-body Hamiltonian we have studied is adiabatically connected to the MR state.

\section{Additional evidence for MR state}

\edit{Here, we provide further evidence for the Moore--Read state at $\nu=1/2$ through the energy and particle-cut entanglement spectra across many system sizes.}
In the main text, we primarily worked with ED clusters of size $N = N_1 \times N_2$, which usually were constructed by taking $N_{1,2}$ unit cells along the primitive lattice vectors $\mathbf{a}_{1,2}$ of the kagome lattice, where $\mathbf{a}_1 = (1, 0),\, \mathbf{a}_2  = \frac{1}{2}(1, \sqrt{3})$.
Therefore, the real-space cluster was spanned by boundary vectors $\mathbf{L}_1 = N_1 \mathbf{a}_1,\, \mathbf{L}_2 = N_2 \mathbf{a}_2$.
However, it is also possible to instead construct clusters with boundary vectors that are linear combinations of the lattice vectors, \ie $\mathbf{L}_1 = n_{x1} \mathbf{a}_1 + n_{y1} \mathbf{a}_2,\, \mathbf{L}_2 = n_{x2} \mathbf{a}_1 + n_{y2} \mathbf{a}_2$ for cluster integers $n_{x1}, n_{y1}, n_{x2}, n_{y2}$, which allows for more control over the aspect ratio of the cluster for a given system size.
For a cluster with $N$ unit cells, it can be shown that the four cluster integers must satisfy $|n_{x1} n_{y2} - n_{x2}n_{y1}| = N$ (see \cite{repellin2014z2} for more details).

\begin{figure}
    \centering
    \includegraphics[scale=1]
    {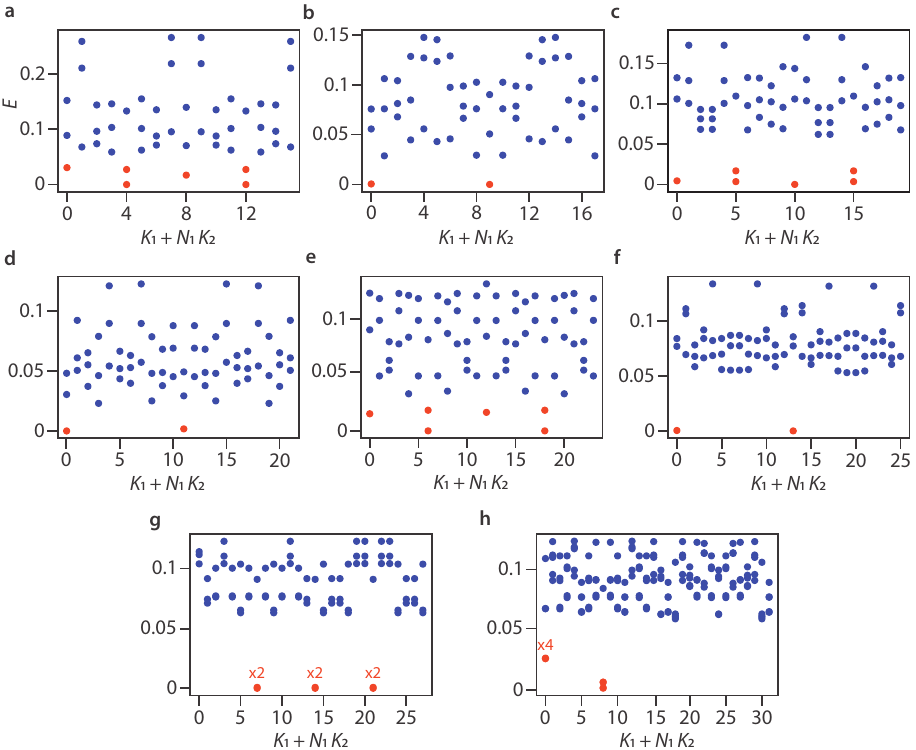}
    \caption{%
    \edit{ED spectra for the non-Abelian MR state in the main text for additional tilted clusters.
    (a)~$N = 16$ cluster with cluster integers $(n_{x1}, n_{y1}, n_{x2}, n_{y2}) = (4, -4, 1, 3)$.
    (b)~$N = 18$ cluster with cluster integers $(3, 2, -3, 4)$.
    (c)~$N = 20$ cluster with cluster integers $(4, -5, -4, 0)$.
    (d)~$N = 22$ cluster with cluster integers $(4, -5, 2, 3)$.
    (e)~$N = 24$ cluster with cluster integers $(4, -5, 4, 1)$.
    (f)~$N = 26$ cluster with cluster integers $(4, -5, 2, 4)$.
    (g)~$N = 28$ cluster with cluster integers $(4, -6, 2, 4)$.
    (h)~$N = 32$ cluster with cluster integers $(6, -4, 2, 4)$.}
    }
    \label{fig:figure6}
\end{figure}

\begin{figure}
    \centering
    \includegraphics[scale=1]
    {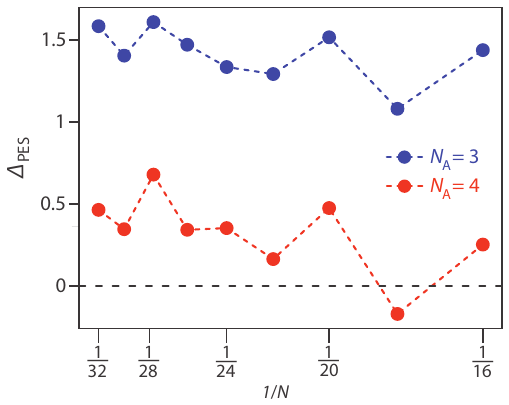}
    \caption{%
    \edit{Particle entanglement spectrum gaps as a function of inverse system size for subspace particle numbers $N_A = 3, 4$.}
    }
    \label{fig:pes_mr}
\end{figure}

In \cref{fig:figure6} we show ED spectra \edit{across several system sizes}, all of which have aspect ratio greater than $0.8$.
We observe that the MR state \edit{is clearly separated from excited states} for these clusters \edit{but with varying spreads}, indicating that the ground state is sensitive to the system's aspect ratio, as discussed in the main text.
\edit{Furthermore, we show in \cref{fig:pes_mr} the particle entanglement spectrum gap $\Delta_{\operatorname{PES}}$ as a function of inverse system size, for the same clusters as in \cref{fig:figure6}. In our definition, $\Delta_{\operatorname{PES}} > 0$ indicates that every momentum sector has the exact count expected for the Moore--Read state \cite{bernevig_emergent_2012}, with a global gap above it. Except for $N = 18, N_A = 4$, every system size shows a positive entanglement gap, with an overall upward trend suggesting that this feature persists in the thermodynamic limit.}

\bibliography{references}